\documentclass{aa}  
\usepackage{natbib}
\bibpunct{(}{)}{;}{a}{}{,}
\usepackage{graphicx}
\usepackage{enumitem}
\usepackage{epstopdf}
\usepackage[percent]{overpic}
\usepackage{comment}
\usepackage{color}
\usepackage{url}
\usepackage{breqn}
\usepackage{xcolor}
\usepackage[backref=page]{hyperref}
\newcommand{\ca}{\mbox{Ca\,{\sc ii}~K\,}}

\begin{document}
\authorrunning{Chatzistergos et al.}
\titlerunning{Plage-sunspot relation}
\title{Scrutinizing the relationship between plage areas and sunspot areas and numbers}
\newcounter{affiliations}
\author{Theodosios~Chatzistergos\inst{\refstepcounter{affiliations}{\theaffiliations},\refstepcounter{affiliations}{\theaffiliations}}, Ilaria~Ermolli\inst{2},  
        Natalie~A.~Krivova\inst{1}, 
        Teresa~Barata\inst{\refstepcounter{affiliations}{\theaffiliations}\label{CESR1}}, 
	Sara~Carvalho\inst{\refstepcounter{affiliations}{\theaffiliations}}, 
	Jean-Marie~Malherbe\inst{\addtocounter{affiliations}{1}\theaffiliations,\addtocounter{affiliations}{1}\theaffiliations}
        }
\offprints{Theodosios Chatzistergos  \email{chatzistergos@mps.mpg.de}}
\institute{ Max Planck Institute for Solar System Research, Justus-von-Liebig-weg 3,	37077 G\"{o}ttingen, Germany 
\and INAF Osservatorio Astronomico di Roma, Via Frascati 33, 00078 Monte Porzio Catone, Italy 
\and Univ Coimbra, Instituto de Astrofísica e Ciências do Espaço, Department of Earth Sciences,  3030-790   Coimbra, Portugal
\and Univ Coimbra, Instituto de Astrofísica e Ciências do Espaço, 3040-004 Coimbra, Portugal
\and LESIA, Observatoire de Paris, 92195 Meudon, France
\and PSL Research University, Paris, France}
\date{}

\abstract
{Studies and reconstructions of past solar activity require data on all magnetic regions on the surface of the Sun, i.e. on dark sunspots as well as bright faculae/plage and network. 
Such data are also important for understanding the magnetic activity and variability of the Sun and Sun-like stars. 
The longest available direct faculae/plage datasets are white-light facular and \ca observations going back to 1874 and 1892, respectively. 
Prior to that time, the only direct data available are for sunspots.}
{We reassess the relationship between plage areas and sunspot records (areas and numbers) since 1892, to allow reconstructions of facular/plage areas which can be employed for studies going further back in time, i.e. over the period when solely sunspot observations are available.} 
{We use the plage areas derived from 38 consistently processed \ca archives as well as the plage area composite based on these archives. 
The considered archives include both the well-known ones (e.g. Coimbra, Kodaikanal, Meudon, Mt Wilson), and less explored ones (such as those from the Kharkiv, Mees, and Upice  observatories). 
These data allow us to study the relationship between plage area and sunspot records (areas and number) over a period of twelve solar cycles and for different bandpasses.}
{We find the relationship between plage and sunspot areas to be well represented by a power law function. 
Also the relationship between the plage areas and the sunspot number is best fit with a power law function.
We further find that the relationships depend on the bandwidth and the solar cycle strength. 
The reconstructions with a power law relationship are in good agreement with the original plage area series, whereas employment of a cycle-strength-dependent relationship improves the reconstructions only marginally.
We also estimate the error in the plage areas reconstructed from the sunspot areas or numbers. 
Performing the same analysis on other previously published plage area series, usually derived from a single archive with diverse processing techniques, returns different results when using different time series. 
This highlights the importance of applying a consistent processing to the various archives and demonstrates the uncertainties introduced by using previously published series for studies of past solar activity, including irradiance reconstructions. }
{Our results have implications for past solar activity and irradiance reconstructions as well as for stellar activity studies, which sometimes assume a linear dependence between plage and sunspot areas.
}
	
\keywords{Sun: activity - Sun: photosphere - Sun: chromosphere - Sun: faculae, plages - sunspots}
	
\maketitle
	
\section{Introduction}
\label{sec:intro}
\sloppy

\begin{table*}
	\caption{Previous studies of the relationship between sunspot and facular measurements.}
	\label{tab:previousstudies}     
	\centering                      
	\begin{tabular}{l*{6}{c}}       
		\hline\hline                
		Study & 			Plage data &Sunspot data 				      		& Period &\multicolumn{3}{c}{Relation}\\
		&&		     		&		 &Annual&Monthly&Daily\\
		\hline                                  
\multicolumn{7}{c}{\ca plage areas and sunspot number series} \\[1mm]
\cite{dorotovic_north-south_2010}&Co&ISNv1&1996--2006&&&quadratic\tablefootmark{a}\\
		\cite{foukal_behavior_1996}		&MW					& ISNv1			&1915--1984&linear			&			&\\
		\cite{fligge_long-term_1998}	&MW\tablefootmark{b}& ISNv1				&1944--1954&		&		&		quadratic \\
		\cite{kuriyan_solar_1982}  		&Ko					& ISNv1			&1905--1977&linear			& 		 	&\\
		\cite{bertello_correlation_2016}&Ko\tablefootmark{c}&ISNv2				&1907--1999&linear			&linear		&		\\ 
		\cite{priyal_long-term_2017}	&Ko					 &ISNv2				&1907--2007&		&linear		&\\
		\cite{singh_determining_2021}	&Ko					 &ISNv2				&1905--2007&		&linear		&linear\\
		\cite{singh_application_2022}	&MW					 &ISNv2				&1915--1985&		&linear		&linear\\
		\cite{singh_application_2022}	&ML					 &ISNv2				&1998--2015&		&linear		&linear\\
\cite{yeo_how_2020}	&SP\tablefootmark{d}&ISNv1, ISNv2		&1976--2017&  &		&power law\tablefootmark{e}\\
\cite{yeo_how_2020}	&SP\tablefootmark{d}&CEA17, HoSc98 		&1976--2017&  &		&power law\tablefootmark{e}\\
\hline\\[-3mm]
\multicolumn{7}{c}{\ca plage areas and sunspot area series} \\[1mm]
		\cite{foukal_effect_1979}		&SGD				& SGD  		&1969--1974& 				& 			&($\sim$0.5--685)\tablefootmark{f}  \\
		\cite{schatten_importance_1985}	&SGD				& SGD 			&1969--1982&				&($\sim$18--40)	&\\ 
		\cite{lawrence_sunspot_1987}	&SGD				& SGD 			&1974--1985&($\sim$5--30)	&			&\\ 
		\cite{foukal_curious_1993}		&SGD				& RGO 			&1954--1987&linear			&			&\\
		\cite{foukal_behavior_1996}		&MW					& RGO 			&1915--1984&linear			&			&\\
		\cite{foukal_what_1998}		    &MW					& RGO 			&1915--1984&linear 		&			&quadratic\\
		\cite{tlatov_new_2009}			&MW					& RGO 			&1944--1963&				&quadratic	&		\\	
		\cite{tlatov_new_2009}			&Ko					& RGO 			&1944--1963&				&quadratic 	&	\\	
		\cite{mandal_association_2017}	&Ko\tablefootmark{g} &RGO 			&1907--1965&linear		&		&\\
		\cite{singh_determining_2021}	&Ko					 &RGO			&1905--2007&		&linear		&linear\\
		\cite{singh_application_2022}	&MW					 &RGO				&1915--1985&		&linear		&linear\\
		\cite{singh_application_2022}	&ML					 &RGO				&1998--2015&		&linear		&linear\\
        \cite{chowdhury_analysis_2022}  &Ko\tablefootmark{g}  &RGO          &1907--1980&quadratic&&\\
		\cite{chapman_solar_1997}		&SF1, SF2					&SF1, SF2 			&1988--1995&quadratic\tablefootmark{h}&quadratic  &		\\ \cite{chapman_improved_2001}	&SF1, SF2					&SF1, SF2 			&1988--1999&				&($\sim$5--40)	    &		\\ 
		\cite{chapman_facular_2011}		&SF1, SF2					& SF1, SF2 			&1988--2009&linear			&linear		&		\\  
		\cite{steinegger_sunspot_1996}  &SP					& VTT\tablefootmark{i} 			&1980	   &				&			& ($\sim$3--30)\\
        \cite{yeo_how_2020}	&SP\tablefootmark{d}&BEA09\tablefootmark{j}	&1976--2017&  &		&power law\tablefootmark{e}\\
        \cite{chatzistergos_historical_2020}	&SP					&BEA09\tablefootmark{j} 		&1960--2002&($\sim$35--55)&		&\\
        \cite{chatzistergos_historical_2020}	&Ky\tablefootmark{k}&BEA09\tablefootmark{j} 		&1928--1969&($\sim$13--20)&		&\\
\hline\\[-3mm]
\multicolumn{7}{c}{Magnetogram facular areas and sunspot area/number series} \\[1mm]
		\cite{solanki_solar_2013}		&KP\tablefootmark{l}& KP areas			&1974--2002&			&			&quadratic	\\ 
		\cite{shapiro_variability_2014}	&KP\tablefootmark{l}& KP areas			&1974--2002&				&			&quadratic	\\
        \cite{borgniet_using_2015}		&MDI\tablefootmark{m}&SOON\tablefootmark{n} areas		&1996--2007&				&($\sim$10--160)	&	\\
        \cite{criscuoli_angular_2016}	&MDI\tablefootmark{m}&ISNv2			&1996--2011&			&linear		&quadratic\\	
		\hline
	\end{tabular}
	\tablefoot{Columns are: the bibliography entry, the type of plage and sunspot data used, the period covered by the analysed data, and the form of the derived relationship \textbf{(considering sunspot data as a function of plage areas} for annual, monthly, and daily data, the values in parentheses are the reported values for the plage to sunspot areas ratio. The column corresponding to the relationship for monthly values is for periods intermediate to annual and daily, while only \cite{tlatov_new_2009}, \cite{bertello_correlation_2016}, and \cite{singh_determining_2021} used monthly values. \cite{chapman_solar_1997,chapman_improved_2001}, and \cite{chapman_facular_2011} used 100-day bins, \cite{shapiro_variability_2014} 58-day bins, \cite{criscuoli_angular_2016} and \cite{priyal_long-term_2017} 6-month averages. See Tables \ref{tab:observatories} and \ref{tab:sunspotseries} as well as footnotes in this table for information on the abbreviations.	
\tablefoottext{a}{This is not explicitly stated by the authors but can be inferred from Fig. 6A by \cite{dorotovic_north-south_2010}.} 
\tablefoottext{b}{The MW plage area series used in this study was the one produced by \cite{foukal_behavior_1996}.} 
\tablefoottext{c}{The Ko plage area series used in this study was the one produced by \cite{tlatov_new_2009}.} 
\tablefoottext{d}{This is the \cite{bertello_correlation_2016} disc-integrated 1\AA~\ca composite index with SP and Synoptic Optical Long-term Investigations of the Sun Integrated Sunlight Spectrometer data. The SP data are not the full-disc spectroheliograms used in our study.}
\tablefoottext{e}{This is a convolution of a power law relation with a finite impulse response filter.}
\tablefoottext{f}{This was roughly estimated from Fig. 4 by \cite{foukal_effect_1979} who argued that there is large scatter in the data rendering sunspot areas a poor indicator of plage areas.}
\tablefoottext{g}{The Ko plage area series used in this study was the one produced by \cite{chatterjee_butterfly_2016}.} 
\tablefoottext{h}{The authors reported a quadratic relation for annual values, however they also state that roughly 75\% of the variance in the scatter plots is accounted by the linear term.}
\tablefoottext{i}{This refers to data acquired with the Vacuum Tower Telescope at the Teide observatory.} 
\tablefoottext{j}{This refers to the \cite{balmaceda_homogeneous_2009} sunspot area composite series.}
\tablefoottext{k}{This refers to the plage areas from the Kyoto observatory.}
\tablefoottext{l}{This refers to facular areas from Kitt Peak magnetograms.} 
\tablefoottext{m}{This refers to facular areas from the Solar and Heliospheric Observatory Michelson Doppler Imager magnetograms.} 
\tablefoottext{n}{This refers to the Solar Optical Observing Network.} 
	}
\end{table*}

\begin{table*}
	\caption{Abbreviations used for \ca plage datasets.} 
	\label{tab:observatories}      
	\centering                                     
	\begin{tabular}{l*{3}{c}}          
		\hline\hline                        
		\textbf{Dataset} & Acronym	&Period   & Reference\\
\hline
\multicolumn{4}{c}{Individual series}\\
Arceti & Ar&1931--1974&\cite{ermolli_digitized_2009},\\
Big Bear&BB&1982--2006&\cite{naqvi_big_2010}\\
Coimbra&Co&1925--2019&\cite{lourenco_solar_2019}\\
Kharkiv&Kh&1952--2019&\cite{belkina_ccd_1996}\\
Kodaikanal&Ko&1904--2007&\cite{priyal_long-term_2017}\\
Mauna Loa\tablefootmark{a}&ML&1998--2015&\cite{rast_latitudinal_2008}\\
McMath-Hulbert&MM&1948--1979&\cite{mohler_mcmath-hulbert_1968}\\
Mees&MS&1982--1998\\
Meudon&MD&1893--2019&\cite{malherbe_new_2019}\\
Mt Wilson&MW&1915--1985&\cite{lefebvre_solar_2005}\\
Rome\tablefootmark{b}&RP&1996--2019&\cite{ermolli_photometric_2007}\\
Sacramento Peak&SP&1960--2002&\cite{tlatov_new_2009}\\
San Fernando CFDT1\tablefootmark{c}&SF1&1988--2015&\cite{chapman_solar_1997}\\
San Fernando CFDT2\tablefootmark{d}&SF2&1992--2013&\cite{chapman_solar_1997}\\
Upice&UP&1998--2019&\cite{klimes_simultaneous_1999} \\
\hline
\multicolumn{4}{c}{Composite series}\\
Solar Geophysical Data series&SGD&1942--1987\\
Plage area composite&CEA20&1892--2019&\cite{chatzistergos_analysis_2020}\\
\hline\end{tabular}
\tablefoot{The \ca plage area series are separated into those resulting from individual archives and composite series.
\tablefoottext{a}{Taken with the Mauna Loa Precision Solar Photometric Telescope, PSPT.}
\tablefoottext{b}{Taken with the Rome Precision Solar Photometric Telescope, PSPT.}
\tablefoottext{c}{Taken with the Cartesian Full-Disc Telescope, CFDT1.}
\tablefoottext{d}{Taken with the Cartesian Full-Disc Telescope, CFDT2.}
}
\end{table*}

\begin{table}
	\caption{Acronyms used in this work to refer to various sunspot datasets, listed separately for the sunspot area, ISN and GSN series. }
	\label{tab:sunspotseries}      
	\centering                                     
	\begin{tabular}{l*{2}{c}}          
		\hline\hline                        
		\textbf{Data series} & Acronym	&Period   \\
\hline\\[-3mm]
\multicolumn{2}{c}{Sunspot areas} \\[1mm]
Royal Greenwich Observatory&RGO&1874--1976\\
\cite{mandal_sunspot_2020}\tablefootmark{a}&MEA20&1874--2019\\
\hline\\[-3mm]
\multicolumn{2}{c}{ISN} \\[1mm]
\cite{clette_wolf_2007}\tablefootmark{b}&ISNv1&1700--2014\\
\cite{clette_new_2016-1}\tablefootmark{b}&ISNv2&1700--2019\\
\hline\\[-3mm]
\multicolumn{2}{c}{GSN} \\[1mm]
\cite{svalgaard_reconstruction_2016}\tablefootmark{b}&SvSc16&1610--2015\\
\cite{chatzistergos_new_2017}\tablefootmark{a}&CEA17&1739--2010\\
\hline\end{tabular}
\tablefoot{
\tablefoottext{a}{The series is available at \url{http://www2.mps.mpg.de/projects/sun-climate/data.html}}
\tablefoottext{b}{Available at \url{https://wwwbis.sidc.be/silso/datafiles}}
}\end{table}

\begin{table*}
	\caption{Best fit parameters between the daily values of CEA20 plage area composite and various sunspot series.}      
	\label{tab:sgnfitparameters}     
	\centering                       
	\begin{tabular}{l*{4}{c}}        
		\hline\hline                 
&\multicolumn{3}{c}{$p_a=a_1+a_2\times s^{a_3}$}&\\
&$a_1$&$a_2$&$a_3$&RSS/DOF\\
\hline
\multicolumn{5}{c}{Power law}\\
\hline
MEA20&-0.004$\pm$ 0.008&  0.342$\pm$  0.093& 0.35$\pm$ 0.08&   0.066\\ 
ISNv1& 0.002$\pm$ 0.002&  0.009$\pm$  0.001& 0.66$\pm$ 0.05&   0.116\\
ISNv2& 0.002$\pm$ 0.002&  0.009$\pm$  0.001& 0.69$\pm$ 0.04&   0.074\\ 
CEA17& 0.004$\pm$ 0.001&  0.005$\pm$  0.001& 0.86$\pm$ 0.06&   0.078\\

\hline
\multicolumn{5}{c}{Squared root}\\
\hline
MEA20& 0.006$\pm$ 0.002&  0.606$\pm$  0.028&0.5&   0.110\\
\hline
\multicolumn{5}{c}{Linear}\\
\hline
MEA20& 0.020$\pm$ 0.001&  5.476$\pm$  0.261&1&   0.604\\ 
ISNv1& 0.011$\pm$ 0.001& 0.0027$\pm$ 0.0001&1&   0.349\\
ISNv2& 0.009$\pm$ 0.001& 0.0033$\pm$ 0.0001&1&   0.304\\ 
CEA17& 0.007$\pm$ 0.001& 0.0034$\pm$ 0.0001&1&   0.114\\
\hline\end{tabular}
	\tablefoot{RSS/DOF are the sum of the squared residuals per degree of freedom. The ISNv1 and ISNv2 sunspot number values were divided by 12.08 (ISNv2 values are also multiplied with 0.6) prior to the fit so to bring them roughly to the same level as the group sunspot number series.}
	\end{table*}

\begin{table*}
	\caption{Best fit parameters between the annual median values of CEA20 plage area composite and annual median values of various sunspot series.}      
	\label{tab:sgnfitparametersannual}      
	\centering                                      
	\begin{tabular}{l*{4}{c}}         
		\hline\hline                       
&\multicolumn{3}{c}{$p_a=a_1+a_2\times s^{a_3}$}&\\
&$a_1$&$a_2$&$a_3$&RSS/DOF\\
\hline
\multicolumn{5}{c}{Power law}\\
\hline
MEA20& 0.007$\pm$ 0.001& 0.0176$\pm$ 0.0013& 0.78$\pm$ 0.06& 0.68\\
ISNv1& 0.005$\pm$ 0.001& 0.0049$\pm$ 0.0009& 0.87$\pm$ 0.06& 0.58\\
ISNv2& 0.005$\pm$ 0.001& 0.0048$\pm$ 0.0008& 0.94$\pm$ 0.07& 0.69\\
CEA17& 0.005$\pm$ 0.001& 0.0029$\pm$ 0.0007& 1.12$\pm$ 0.10& 0.66\\
SvSc16& 0.005$\pm$ 0.001& 0.0038$\pm$ 0.0009& 1.03$\pm$ 0.09& 0.67\\
\hline
\multicolumn{5}{c}{Squared root}\\
\hline
MEA20& 0.0018$\pm$ 0.0008& 0.0247$\pm$ 0.0008&0.5& 0.96\\
\hline
\multicolumn{5}{c}{Linear}\\
\hline
MEA20& 0.0090$\pm$ 0.0006& 0.0137$\pm$ 0.0004&1& 0.77\\
ISNv1& 0.000$\pm$ 0.000& 0.0070$\pm$ 0.0007&1& 0.61\\
ISNv2& 0.000$\pm$ 0.000& 0.0062$\pm$ 0.0006&1& 0.70\\
CEA17& 0.000$\pm$ 0.000& 0.0040$\pm$ 0.0007&1& 0.67\\
SvSc16& 0.000$\pm$ 0.000& 0.0044$\pm$ 0.0007&1&0.67\\
\hline\end{tabular}
	\tablefoot{RSS/DOF are the sum of the squared residuals per degree of freedom. The ISNv1 and ISNv2 sunspot number values were divided by 12.08 (ISNv2 values are also multiplied with 0.6) prior to the fit so to bring them roughly to the same level as the group sunspot number series.}
	\end{table*}

\begin{figure*}[t]
	\centering
\includegraphics[width=0.77\linewidth]{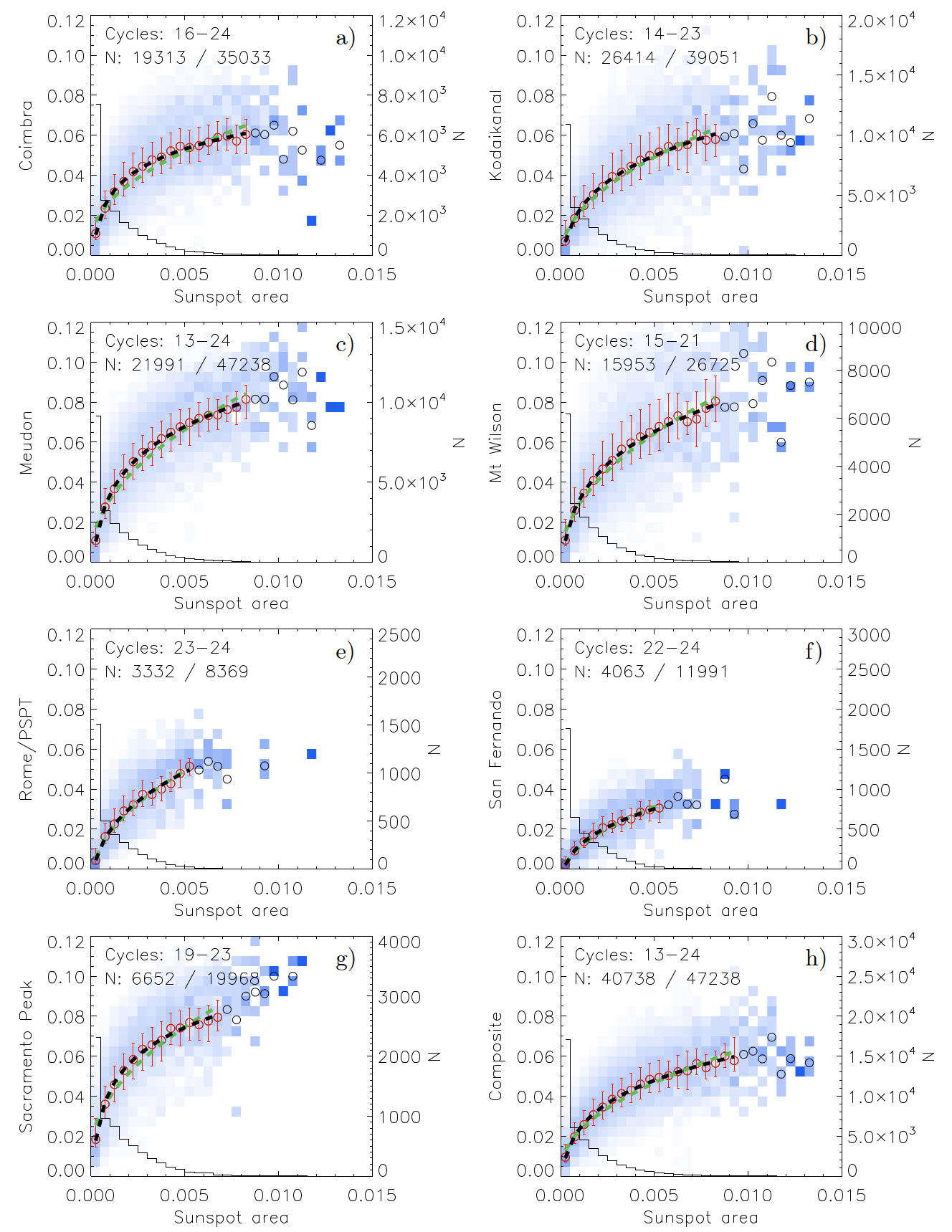}
	\caption{Probability distribution functions (PDF) for plage fractional areas as a function of the sunspot fractional areas from MEA20. Both plage and spot areas were taken on the same days. The PDFs are shown in bins of 0.005 and 0.0005 fractional areas for plage and sunspots, respectively. Each panel shows the PDF matrix for a given Ca II K archive; Co (a)), Ko (b)), MD (c)), MW (d)), RP (e)), SF2 (f)), SP (g)), and the CEA20 plage area composite (h)).  The PDFs are colour-coded between white for 0 and bright blue for 1. 
	Circles denote the average plage area value within each column. 
	For columns with less than 20 days of data these circles are shown in black, otherwise in red. We also show the asymmetric 1$\sigma$ interval for each column. Two different fits to the average values are over-plotted, a power law fit (black), and a square root function (green). 
	The number of days included in each column is also shown as a solid black histogram (see right-hand axis). The period of overlap between the two archives shown in a given panel (expressed in solar cycles), as well as the total number of overlapping days used to construct each matrix (N) and the total number of days within this time interval are listed in the top part of each panel.  }
	\label{fig:pdf_indarch_allperiod}
\end{figure*}

\begin{figure*}[t]
	\centering
\includegraphics[width=0.8\linewidth]{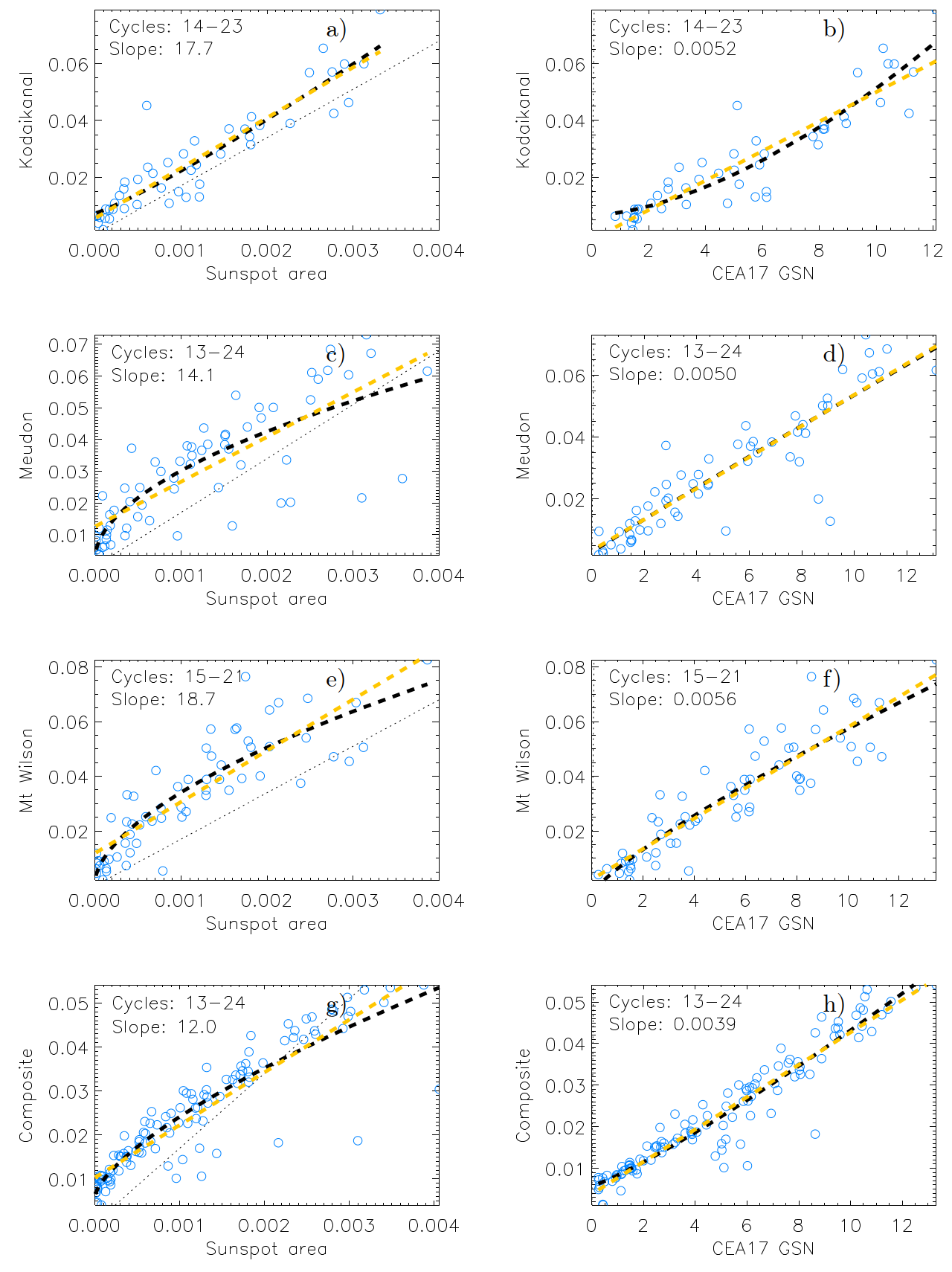}
	\caption{Scatter plots between annual plage fractional areas and the annual MEA20 sunspot fractional areas (left columns) as well as the annual CEA17 GSN series (right columns).
Two different fits are over-plotted, a power \textbf{law (black) and a linear (yellow) fit}. The dotted line has a slope of 17, which corresponds to the mean ratio between the CEA20 plage area series and the MEA19 sunspot area series (see Sect. \ref{sec:relations}).
The period of overlap between the two archives  (expressed in solar cycles) as well as the slope of the linear fit are shown in each panel.}
	\label{fig:pdf_indarch_allperiod_annual}
\end{figure*}

\begin{figure*}[t]
	\centering
	\begin{minipage}{0.84\textwidth}
\includegraphics[width=1\linewidth]{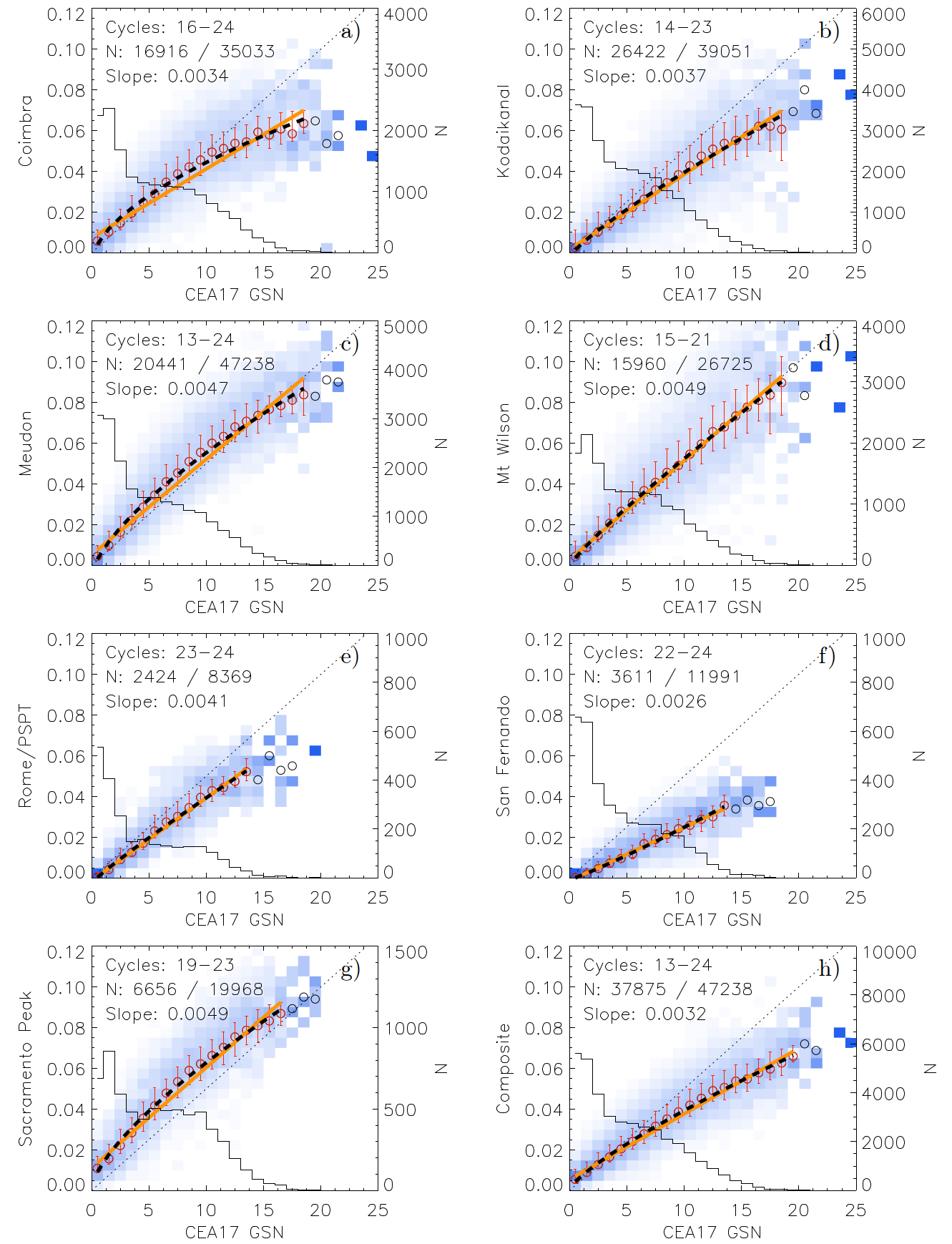}
	\end{minipage}
	\caption{Same as Fig. \ref{fig:pdf_indarch_allperiod} but for the CEA17 GSN series. A linear fit (orange) and a power law fit (black dashed) were performed to the mean values of each column. The slope of the linear fit is also listed in each panel. The black dotted line has a slope of 0.005.}
	\label{fig:pdf_indarch_allperiod_cea17}
\end{figure*}

\begin{figure*}
	\centering
\includegraphics[width=1\linewidth]{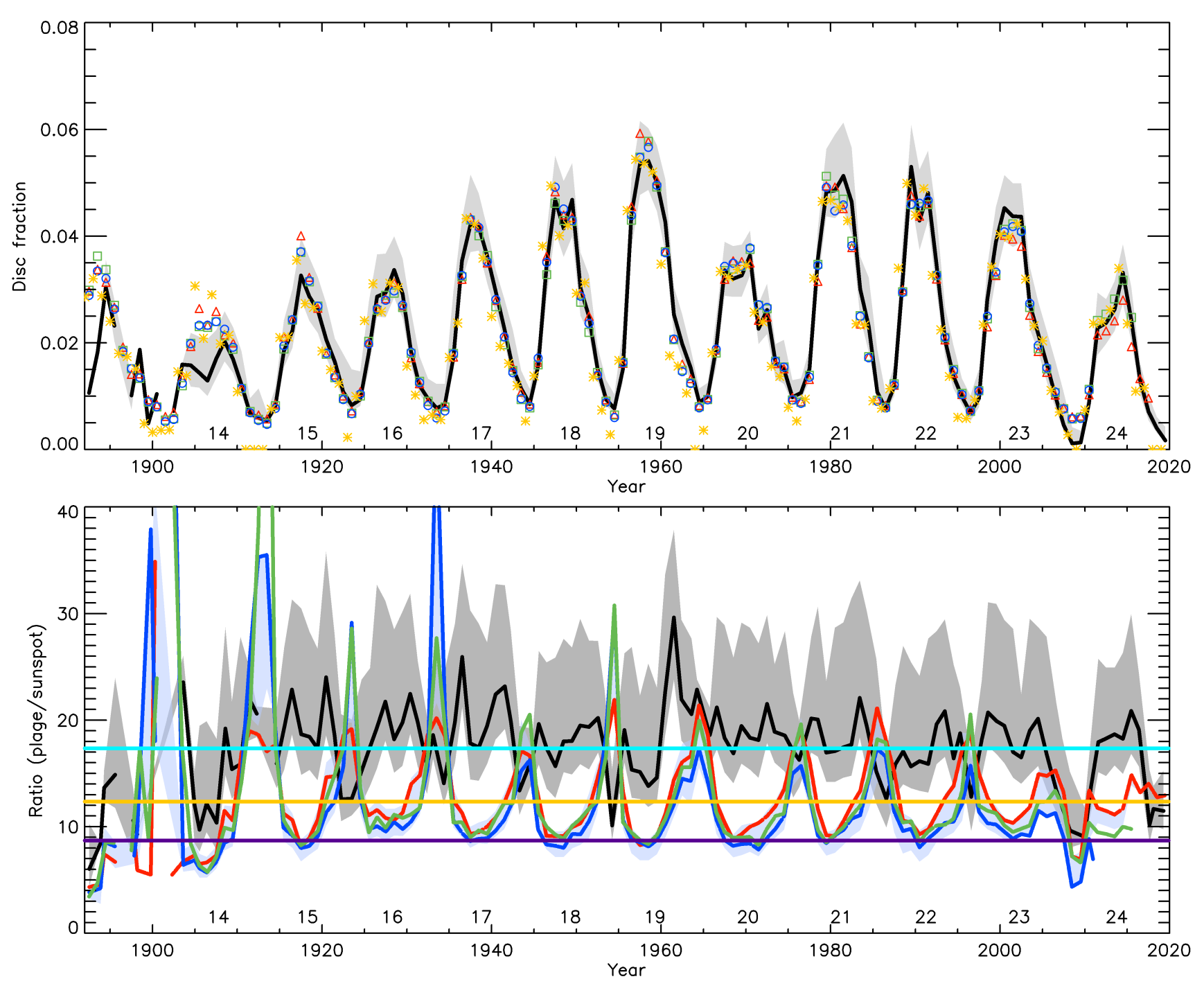}
	\caption{\textit{Upper panel:} CEA20 plage area composite (black line) along with scaled sunspot series by MEA20 (yellow asterisks), CEA17 (blue circles), ISNv2 (red triangles), and SvSc16 (green squares). All series are scaled linearly to the plage areas except MEA20 which is scaled with a square root function. The solid lines are annual median values while the grey shaded surface is the asymmetric 1$\sigma$  interval. \textit{Lower panel:} Ratio of the CEA20 plage areas composite to the sunspot area series by MEA20 (black) and sunspot number series by CEA17 (blue), SvSc16 (green), and ISNv2 (red) as a function of time. The ratios to the GSN series have been multiplied by 2000 to allow a direct comparison to the ratios to the sunspot areas. The blue shaded surface marks the range of annual values for the plage areas to sunspot number ratios from all sunspot number series used in this study. The ratios are calculated only for the days on which sunspot areas are greater than 0.0005 of the disc area or the sunspot number is greater than 0. The yellow (light blue) horizontal line marks the mean ratio of plage areas to sunspot numbers (sunspot areas), while the purple horizontal line is the mean ratio for the plage areas to sunspot number series over four-year intervals around cycle maxima.
	The numbers below the curves in each panel denote the conventional SC numbering and are placed at times of SC maxima.}
	\label{fig:ratiomwsunspotnumberseries}
\end{figure*}

Sunspots and faculae are the most prominent manifestations of solar surface magnetism \citep{solanki_solar_2006-1}.
Sunspots are relatively dark and cool areas on the surface of the Sun, whereas faculae are comparatively small and patchy bright regions usually seen in the vicinity of sunspots and in remnants of active magnetic regions.  
Faculae were originally discovered in full-disc white light images, where they are seen mostly near the limb. 
The co-spatial chromospheric features observed in the \ca spectral range, called plage, are bright and easily observable over the whole solar disc \citep{solanki_smallscale_1993}.
The connection of facular regions to strong magnetic fields was first pointed out by \cite{babcock_suns_1955}.
Being a manifestation of the same physical process, the concentration of magnetic field at the solar surface, sunspots and faculae are closely associated \citep{solanki_sunspots_2003}.
However, their evolution differs.
While individual facular elements generally do not live as long as sunspots, within a given active region facular regions can last much longer than sunspots, partly because sunspot decay products form new facular elements \citep[e.g.][]{wang_rapid_2012}.

Sunspots have been observed on the solar disc since antiquity \citep{vaquero_sun_2009,arlt_historical_2020}. 
However, regular monitoring of sunspots began only with the advent of the telescope in the early 17th century.
Later, \cite{wolf_mittheilungen_1850} started compiling sunspot measurements by various observers and created the first sunspot number series, the Wolf number, later renamed to international sunspot number (ISNv1, hereafter), which is still produced to this day \citep{clette_wolf_2007}. 
This series goes back to 1818 with daily cadence, while annual values are available back to 1700.
More recently, \cite{hoyt_group_1998} introduced a different measure of solar activity based on sunspot counts, namely the number of sunspot groups or group sunspot number (GSN, hereafter). 
This enabled the early sunspot observations performed between 1609 and 1700 to be exploited as well. 
In the last few years, the database of sunspot data has been scrutinised and updated to include more data \citep[e.g.][]{vaquero_revised_2016,carrasco_sunspot_2018,carrasco_sunspot_2020,carrasco_note_2021,hayakawa_thaddaus_2020,hayakawa_sunspot_2020,hayakawa_daniel_2021,vokhmyanin_sunspot_2020}. 
Also, the method on how to include individual sunspot series has been a matter of debate \citep{lockwood_assessment_2016,usoskin_dependence_2016}. 
This led to the release of a new version of ISNv2 \citep{clette_new_2016-1} and a number of alternative GSN series \citep[e.g.][]{lockwood_centennial_2014,cliver_discontinuity_2016,svalgaard_reconstruction_2016,usoskin_new_2016,usoskin_robustness_2021,chatzistergos_new_2017,willamo_updated_2017}.
In addition to sunspot and sunspot group numbers, the areas of sunspots have also been recorded since the early telescopic observations \citep{arlt_sunspot_2013,arlt_historical_2020}, albeit considerably less systematically than sunspot numbers, which makes a cross-calibration of such individual records extremely challenging. 
Therefore, the earliest sunspot area measurements typically employed in solar activity and irradiance studies are those from the Royal Greenwich Observatory (RGO, hereafter) dating back to 1874.

Observations of bright faculae have also been reported since the advent of telescopes \citep{vaquero_sun_2009}. 
However, due to their low contrast when observed in the continuum, observations of facular regions had long been rather episodic, with regular observations over a long timespan potentially carrying information on such regions going back to the late 19th century.
More detailed information on facular regions is provided by \ca observations, which have been performed at many places around the globe since 1892 \citep[see][and references therein]{chatzistergos_analysis_2017,chatzistergos_analysis_2020} and thus provide a very good temporal coverage of the entire 20th century. 
These observations sample the lower chromosphere, where faculae are seen as bright plage regions. 
Thus, an accurate measurement of plage properties in these observations is of primary interest for reconstructions of past solar magnetism \citep{chatzistergos_recovering_2019}. 

Various studies require information about both sunspots and faculae, such as irradiance reconstructions \citep[e.g.][]{foukal_empirical_1990,solanki_solar_1998,fligge_solar_2000-1,krivova_reconstruction_2010,dasi-espuig_reconstruction_2016,wu_solar_2018-1,lean_estimating_2018}, reconstructions of the long-term evolution of the solar magnetic field \citep[e.g.][]{solanki_evolution_2000,solanki_search_2002,cameron_surface_2010,jiang_solar_2011-1,jiang_modeling_2013,nagovitsyn_area_2016,bhowmik_prediction_2018,krivova_modelling_2021}, or analyses of large-scale flow patterns \citep[][and references therein]{rincon_suns_2018}.

However, even if the \ca data are used directly, they cover only about 130 years, compared with 400 years of sunspot number data. 
The \ca data series are too short for many purposes. 
For example, to get a good idea of solar influence on climate, it is important to have solar irradiance reconstructions going back to the Maunder minimum, that is until at least 1700 CE. 
To do this, it is necessary to reconstruct the plage area starting from sunspot data. 
A first step in that direction is to obtain a reliable relationship between faculae and sunspots spanning more than the last couple of sunspot cycles.

Multiple studies have endeavoured to determine the relationship between faculae and sunspots \citep[e.g.][]{foukal_effect_1979,foukal_behavior_1996,foukal_what_1998,kuriyan_solar_1982,lawrence_sunspot_1987,chapman_solar_1997,tlatov_new_2009,solanki_solar_2013,shapiro_variability_2014,bertello_correlation_2016,criscuoli_angular_2016,yeo_how_2020,de_paula_evolution_2020, berrilli_long-term_2020,nemec_faculae_2022}, by using different data and methodologies. 
Table \ref{tab:previousstudies} summarises the findings of some previous such works, emphasising the analyses that used \ca plage areas as the facular data. 
The majority of the previous studies supported a linear and a quadratic relation (considering sunspot data as a function of plage areas) when annual and daily areas were considered, respectively. 
\cite{lawrence_sunspot_1987,lawrence_ratio_1987} also found that the ratio of the annual plage to sunspot areas varies over the solar cycle (SC, hereafter) being lowest during SC maxima, hence hinting at a non-linear relation.
Furthermore, \cite{lawrence_sunspot_1987}, \cite{chapman_solar_1997}, and \cite{bertello_correlation_2016} reported that the relation depends on the SC strength.
Most of these studies used plage areas derived from uncalibrated historical \ca data \citep[except for][who used CCD-based data]{chapman_solar_1997,chapman_improved_2001,chapman_facular_2011} and mainly either data from a single archive or simply appending data from different observatories without cross-calibrating them, as for example in the Solar Geophysical data series (SGD, hereafter) (see Sect. \ref{sec:data}). 
At the same time, relations derived from historical archives carry significant uncertainties due to intrinsic inconsistencies of the various datasets and their processing (see Sect.~\ref{sec:appendixcomparisonotherplageseries}).
Therefore, any study based on these data, e.g. reconstruction of solar irradiance variations, is also affected by the intrinsic inconsistency of the various archives.  
These limitations can be overcome by exploiting the results of the analysis of \ca observations by \citet{chatzistergos_analysis_2018,chatzistergos_analysis_2019,chatzistergos_analysis_2020}. 

Here we use the consistent record of plage areas derived by \citet[][]{chatzistergos_analysis_2020} from the analysis of 38 \ca archives together with several available series of sunspot observations to study the relationship between plage areas and sunspot data.
We examine not only the functional form of the relationship for various sunspot datasets, but also analyse the dependence of such a relationship on the activity level and the bandwidth of the \ca observations.

In Sect. \ref{sec:data} we give an overview of the various full-disc \ca observations as well as the published plage and sunspot time-series considered in this study. 
In Sect. \ref{sec:comparisonplagesunspots} we study the relationship between the plage areas and the various sunspot records. 
We then use these relationships to reconstruct plage time series based on sunspot observations, which we test by comparing with the actual plage datasets.
We also discuss our findings and compare them to results presented in the literature. 
Finally, we summarise our results and draw our conclusions in Sect. \ref{sec:conclusions}. 

\section{Data}
\label{sec:data}
\subsection{\ca plage series}
We use the time series of projected plage areas produced and made available by \citet[][]{chatzistergos_analysis_2020}.
These plage area series were produced after accurately processing over 290,000 images from 43 historical and modern \ca archives \citep{chatzistergos_exploiting_2016,chatzistergos_ca_2018,chatzistergos_historical_2019,chatzistergos_analysis_2019,chatzistergos_delving_2019,chatzistergos_modelling_2020,chatzistergos_historical_2020,chatzistergos_analysis_2020} spanning the period 1892--2019. 
A composite of plage areas was produced by combining results from 38 archives \citep[][hereafter referred to as CEA20]{chatzistergos_analysis_2020,dataplagecomposite}\footnote{The series is available at \url{http://www2.mps.mpg.de/projects/sun-climate/data.html}} covering the period 1892--2019.
The composite includes separately projected and corrected areas for foreshortening, however here we used only the projected areas. 
The five archives considered by \cite{chatzistergos_analysis_2020}, but finally not used for the composite included observations taken off-band.
While we have actually considered all individual series from \cite{chatzistergos_analysis_2020}, some of them are quite short and their analysis does not fortify the results or conclusions presented in the following.
Therefore, for the sake of simplicity and brevity, we do not show the results for such series here. 
In the following we will present results only for 15 archives summarised in Table \ref{tab:observatories}, where we also give the acronym with which we will refer to them. 
We refer the reader to Table 1 in \cite{chatzistergos_analysis_2020} for the main characteristics of these archives.
One exception is the plage area record from Meudon \citep{malherbe_monitoring_2022}. 
Instead of using the series by \cite{chatzistergos_analysis_2020} we produced a new plage area series by applying exactly the same processing as \cite{chatzistergos_analysis_2020}.
This was done as to include newly digitised data covering mainly the period 1948--1961. 
The new dataset includes 5865 new images, which were missing in the series produced by \cite{chatzistergos_analysis_2020}.

Details of the processing procedure and the derivation of plage areas can be found in \cite{chatzistergos_analysis_2018,chatzistergos_analysis_2019,chatzistergos_analysis_2020}. 
Briefly, images from the photographic archives were photometrically calibrated and compensated for the limb-darkening as described by \cite{chatzistergos_analysis_2018}. 
Also images taken with a CCD were processed to compensate for the limb-darkening with the same method.
All images were segmented to identify plage areas with a multiplicative factor to the standard deviation of the quiet Sun intensity values \citep{chatzistergos_analysis_2019}.
\citet{chatzistergos_analysis_2018,chatzistergos_analysis_2019} showed that the developed method significantly reduces errors in the estimates of plage areas from the various historical \ca archives.

We also considered the previously published \ca plage area series by \cite{kuriyan_long-term_1983}, \cite{foukal_behavior_1996,foukal_comparison_2002}, \cite{ermolli_comparison_2009}, \cite{tlatov_new_2009}, \cite{chatterjee_butterfly_2016}, \cite{priyal_long-term_2017}, \cite{singh_variations_2018}, SGD\footnote{Available at \url{https://www.ngdc.noaa.gov/stp/solar/calciumplages.html}}, as well as the \ca emission index by \cite{bertello_mount_2010}.

The processing and data selection vary considerably among the above published series \citep[see][for more details]{chatzistergos_analysis_2019}.
For example, the series by \cite{kuriyan_long-term_1983} was created by manually selecting the plage regions from the actual Ko photographs.
Similarly, the SGD series includes areas determined manually from the physical photographs of MM (06/1942--09/1979), MW (10/1979--09/1981), and BB (10/1981--11/1987) observatories.
\cite{foukal_behavior_1996} manually selected plage areas from an earlier version of MW dataset digitised with an 8bit device.
\cite{foukal_comparison_2002} derived plage (including the enhanced network) areas from the 8bit MW dataset and extended it with SP data up to 1999. 
\cite{ermolli_comparison_2009} processed the 8bit Ko, and 16bit MW datasets.
\cite{tlatov_new_2009} processed the 8bit Ko data and the 16bit MW data.
\cite{bertello_mount_2010} analysed the 16bit MW data and derived a \ca index, which they found to be related to the plage areas by \cite{tlatov_new_2009}. 
\cite{chatterjee_butterfly_2016}, \cite{priyal_long-term_2017}, and \cite{singh_variations_2018} processed the 16bit digitised Ko archive.
\cite{ermolli_comparison_2009}, \cite{tlatov_new_2009}, \cite{priyal_long-term_2017}, and \cite{singh_variations_2018} applied simple photometric calibration to the data, while SGD, \cite{foukal_behavior_1996,foukal_comparison_2002}, \cite{bertello_mount_2010}, and \cite{chatterjee_butterfly_2016} used photometrically uncalibrated observations.
All the above series are given at a daily cadence except those by \cite{kuriyan_long-term_1983}, \cite{tlatov_new_2009}, and \cite{bertello_mount_2010} which are available as annual values, and the series by \cite{foukal_comparison_2002}, for which 12-month running mean values are provided.
In this work we use the series by \cite{bertello_mount_2010} after applying a linear scaling to match our derived plage area series from the 16bit MW data.
We note that all these series include areas corrected for foreshortening in fractions of the hemisphere, while the ones by \cite{chatterjee_butterfly_2016}, \cite{priyal_long-term_2017}, and \cite{singh_variations_2018} include projected areas.

\subsection{Sunspot series}
For our study we used the sunspot area record compiled by \citet[][MEA20, hereafter]{datasunspotareasmandal}\footnotemark[1]. 
We used the projected areas from this dataset like for the plage ones.
This series is a composite of the RGO (1874--1976), Debrecen (1976--2019), and Kislovodsk (1977--2019) sunspot area records. 

We also used the various available sunspot number and group number series.
These series and  their acronyms used here are summarised in Table \ref{tab:sunspotseries}.
Briefly, ISNv1, ISNv2, and SvSc16 used simple linear scaling to calibrate the records by the various observers.
CEA17 used a non-linear and non-parametric approach to cross-calibrate the counts by different observers, thus taking their diverse observing capabilities into account.

We note that the values of ISNv1 and ISNv2 were divided by 12.08 and 20.13 (12.08/0.6), respectively, to bring them to the level of the GSN series.
Owing to the corrections introduced to ISNv1 to produce ISNv2 \citep[see e.g.,][]{clette_new_2018}, this scaling results in a small discontinuity between ISNv1 and ISNv2 over 1947 \citep{clette_new_2016-1}, with ISNv2 being consistently lower than ISNv1 afterwards.
We further note, that the series by SvSc16 has annual cadence, while the other series have daily values. 

\section{Results}
\label{sec:comparisonplagesunspots}
\subsection{Relationship between plage and sunspot series}
\label{sec:relations}
We first analysed the relationship between the daily plage areas derived from the analysis of each individual \ca archive and the sunspot area series from MEA20 by considering their entire respective overlap periods. 
Following \cite{chatzistergos_new_2017}, we did this by using probability distribution function (PDF) matrices between the daily plage and sunspot areas.
Briefly, for a given archive, the PDF matrices are created by using another archive as the reference.
For this, we first sort the values of the considered archive in bins of specific width. 
For each bin, we compute the histogram of the co-temporal observations from the reference archive.
Each histogram is then normalised to the total number of data points included, thus resulting in a PDF.
Examples of such PDF matrices when considering the \ca series as the references (ordinate) are shown in Fig.~\ref{fig:pdf_indarch_allperiod}.
We note that the PDF matrices were computed with bin sizes of $10^{-3}$, $10^{-4}$, and 0.1, for plage area, sunspot area, and sunspot number series, respectively, however for aiding the visualisation of the PDF matrices in the following they are shown with bins of 5$\times10^{-3}$, 5$\times10^{-4}$, and 1.
We then fit the plage area values averaged over every sunspot-area bin with a square-root and a power law function. 
We note that since our purpose is to use sunspot data to reconstruct plage areas, we expressed plage areas as a function of sunspot data, which is the opposite of how previous results are listed in Table \ref{tab:previousstudies}.
The results are qualitatively similar for all  archives, and the overall shape of the relationship is in general in agreement with previous studies. 
However, we find the power law to fit the data even better than the square-root function reported in previous studies, although at the cost of an additional free parameter (the exponent). 
This also holds when the sunspot areas are taken as the reference.
Considering the CEA20 plage area composite and the MEA20 sunspot area record, we find
the following relationship between them:
\begin{equation}
    p_a=(0.34\pm0.09)\times s_a^{(0.35\pm0.08)}-(0.004\pm0.008),
    \label{eq:plf1}
\end{equation}
where $p_a$ and $s_a$ are the plage and sunspot areas, respectively.

Table \ref{tab:sgnfitparameters} lists the best fit parameters for the fits between the CEA20 plage area and MEA20 sunspot area composite series along with the sums of squared residuals per degree of freedom (RSS/DOF). 
Interestingly, even when considering annual values (seen in Fig. \ref{fig:pdf_indarch_allperiod_annual}), we see a clear tendency for a non-linearity in the relationship, although there is a considerable scatter for some series.
The parameters of the fits for the annual values of the CEA20 plage area composite and the MEA20 sunspot area series are listed in Table \ref{tab:sgnfitparametersannual}.

These results are in agreement with those by \citet{chapman_solar_1997}, but in contrast to those of \cite{foukal_behavior_1996,foukal_what_1998} and \cite{chapman_facular_2011}, who reported a linear relationship between sunspot and plage areas when annual values were considered.

Figure \ref{fig:pdf_indarch_allperiod_cea17} shows the relationship between the various \ca plage area series and the CEA17 GSN series.
To a good first approximation, the relation can be considered linear for all archives, although some archives hint at a weak non-linearity, see e.g. the results for SP, MD, or Co data.
The parameters of the power-law fit between the CEA20 plage area composite and the various sunspot series are given in Table \ref{tab:sgnfitparameters} along with the parameters of a linear fit and the resulting RSS/DOF. 
For the CEA20 plage area composite and the CEA17 GSN series, we find the power law function to fit the data better than the linear relation.

Table \ref{tab:sgnfitparameters} shows that the relation is only mildly affected by the choice of the sunspot series.
For the CEA17 GSN series we adopt the following relationship:  
\begin{equation}
    p_a=(0.005\pm0.001)\times s_{gn}^{(0.86\pm0.06)}+(0.004\pm0.001),
    \label{eq:lnf2}
\end{equation}
where $s_{gn}$ is the group count.

The relation is to a good approximation linear for the annual values (see Fig.~\ref{fig:pdf_indarch_allperiod_annual}h).
However, the slope of the fit to annual values tends to be slightly higher than that to the daily values.  
This has most likely to do with the greater lifetime of plage ensemble regions compared to sunspots.
The parameters of the fits for the annual values of the CEA20 plage area composite and the various sunspot series are listed in Table \ref{tab:sgnfitparametersannual}.
This is in agreement with the results by \cite{kuriyan_solar_1982}, \cite{foukal_behavior_1996}, and \cite{bertello_correlation_2016}, who found a linear relation between annual ISNv1 and \ca plage areas.
We note that the shape of the relationships we find for plage-sunspot areas and plage-sunspot number series imply a non linear relation between sunspot areas and sunspot number series, which is in agreement with previous results, e.g. \cite{fligge_inter-cycle_1997,balmaceda_homogeneous_2009,carrasco_normalized_2016,nagovitsyn_average_2021,mandal_size_2021}.

Figure \ref{fig:ratiomwsunspotnumberseries} (top panel) shows the annual values of the plage area composite along with the scaled series by MEA20, CEA17, SvSc16, and ISNv2. 
We find good agreement between all the series and the CEA20 plage area composite.
An exception is the period before 1905 over which, however, the coverage of the \ca plage area composite is poorer than at other times, which leads to poorer statistics when comparing to sunspot observations.

Figure \ref{fig:ratiomwsunspotnumberseries} (bottom panel) shows the ratio of the plage area composite to the various sunspot series. 
To compute the ratios, we ignored days on which the sunspot number was 0 or sunspot areas were lower than 0.0005 of the disc area. 
We note that a lower threshold for the sunspot areas would leave unaffected the ratio over activity maximum periods, but increase the ratio during activity minima.
The ratio of the plage to sunspot areas is in the range [6,30] with an average of $17\pm4$ when annual values are considered. 
The range is [0.6,147] for the daily values with an average of $21\pm11$. 
These values are consistent with those from \cite{schatten_importance_1985,lawrence_ratio_1987,chapman_improved_2001}, but they are lower than those reported by \cite{chapman_facular_2011}. 
It is noteworthy that, the ratio increases during the descending phase of SC 19.
Also interesting, the ratio of the plage areas to the sunspot number was roughly the same ($\sim$0.0043) during all activity maxima over SCs 15--23.

A careful comparison of different panels of Figs.~\ref{fig:pdf_indarch_allperiod} and \ref{fig:pdf_indarch_allperiod_cea17} seems to hint at changes in the relationship between faculae and spots with time.
For example, panels e and f, which are limited to cycles 22--24 exhibit weaker slopes than relationships covering earlier periods.
In Appendix~\ref{sec:dependenceoncyclestrength} we take a closer look at this and, indeed, find a weak dependence of the relationships on cycle strength.

\subsection{Reconstructing plage areas from sunspot series}
\label{sec:plagefromsunspot}

\begin{figure*}
	\centering
\includegraphics[width=1\linewidth]{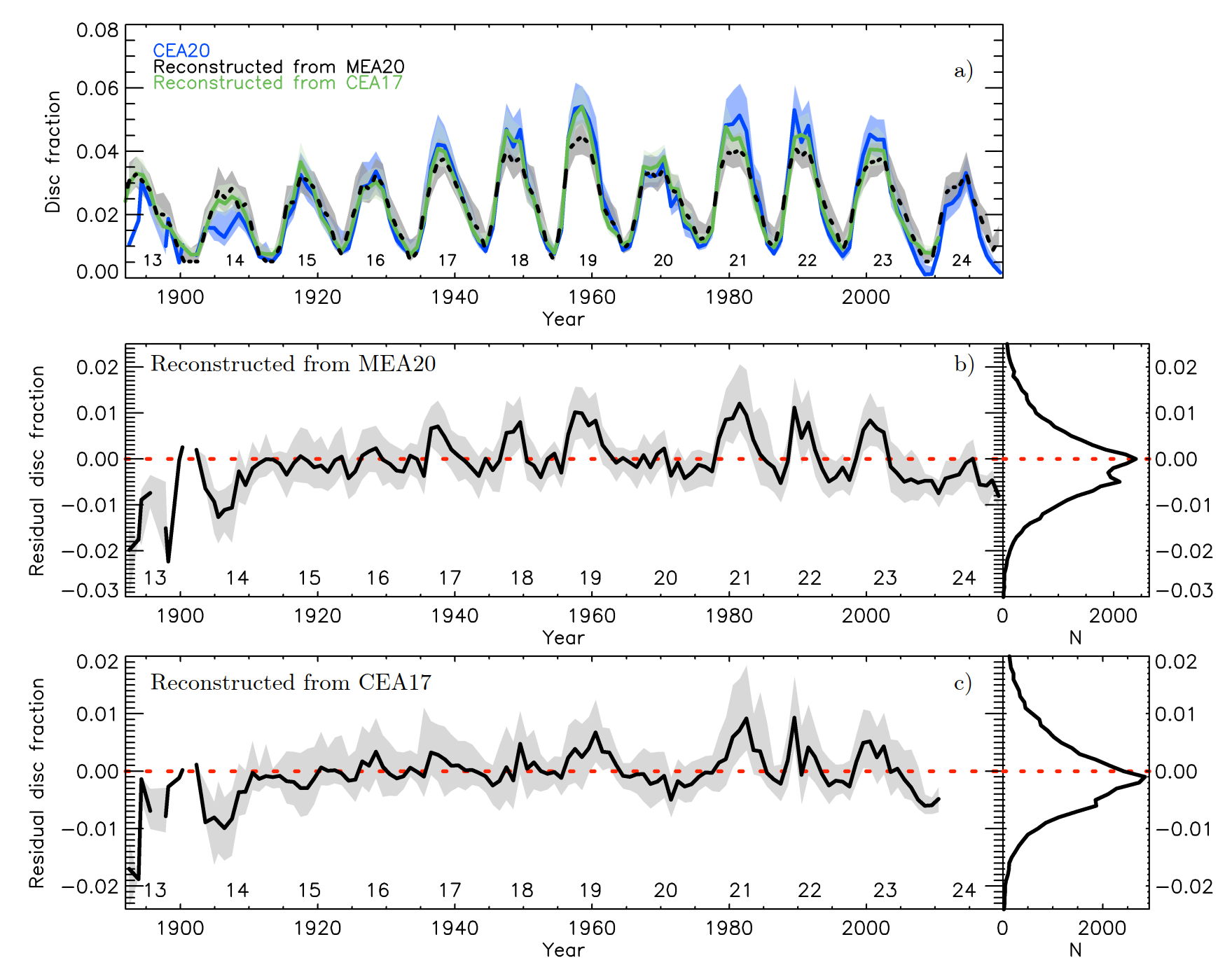}
	\caption{\textit{Panel a: } Reconstructed plage areas from MEA20 sunspot area series (black) and the CEA17 GSN series (green) along with the CEA20 plage area composite (blue). 	Shown are annual median values (solid lines) along with the asymmetric 1$\sigma$ intervals (shaded surfaces) only for the period covered by CEA20 plage area composite. The numbers below the curves denote the conventional SC numbering and are placed at SC maximum periods. 
	\textit{Panels b--c Left: }Residual areas in fractions of the disc between the CEA20 plage area composite and the plage areas reconstructed from the MEA20 sunspot areas (panel b) and the CEA17 GSN series (panel c). 
	The red dashed line denotes 0 plage area differences. \textit{Panels b--c Right: } Distributions of the differences in bins of 0.001 in fractional areas.}
	\label{fig:residualreconstructedplageseriesfromsunspotnumberseriesmain}
\end{figure*}

We now use the relationships derived in Sect.~\ref{sec:relations} to reconstruct plage areas from sunspot data and analyse the performance of these reconstructions. We use the power law function 
on sunspot area (Eq. \ref{eq:plf1}) and GSN (Eq.~\ref{eq:lnf2}), respectively. 
Figure \ref{fig:residualreconstructedplageseriesfromsunspotnumberseriesmain}a shows the reconstructed plage areas from the MEA20 sunspot area and the CEA17 GSN series, along with the CEA20 plage area composite. 
The residual plage areas between the CEA20 composite and the reconstructions from the MEA20 sunspot area and CEA17 GSN series are shown in Fig. \ref{fig:residualreconstructedplageseriesfromsunspotnumberseriesmain}b and c, respectively.
Both reconstructions perform well, with RMS differences (linear correlation coefficients) between the CEA20 plage area composite and the reconstructions from the MEA sunspot area and the CEA17 GSN series being 0.009 and 0.0079 (0.83 and 0.88), respectively.
Appendix \ref{sec:plagefromsunspot_appendix} presents also reconstructions using a linear, quadratic, or power law relation with SC-dependent parameters, as well as using ISNv2 for reconstructing plage areas.

\subsection{Comparison to other published plage areas}
\label{sec:appendixcomparisonotherplageseries}

\begin{figure*}
	\centering
\includegraphics[width=1\linewidth]{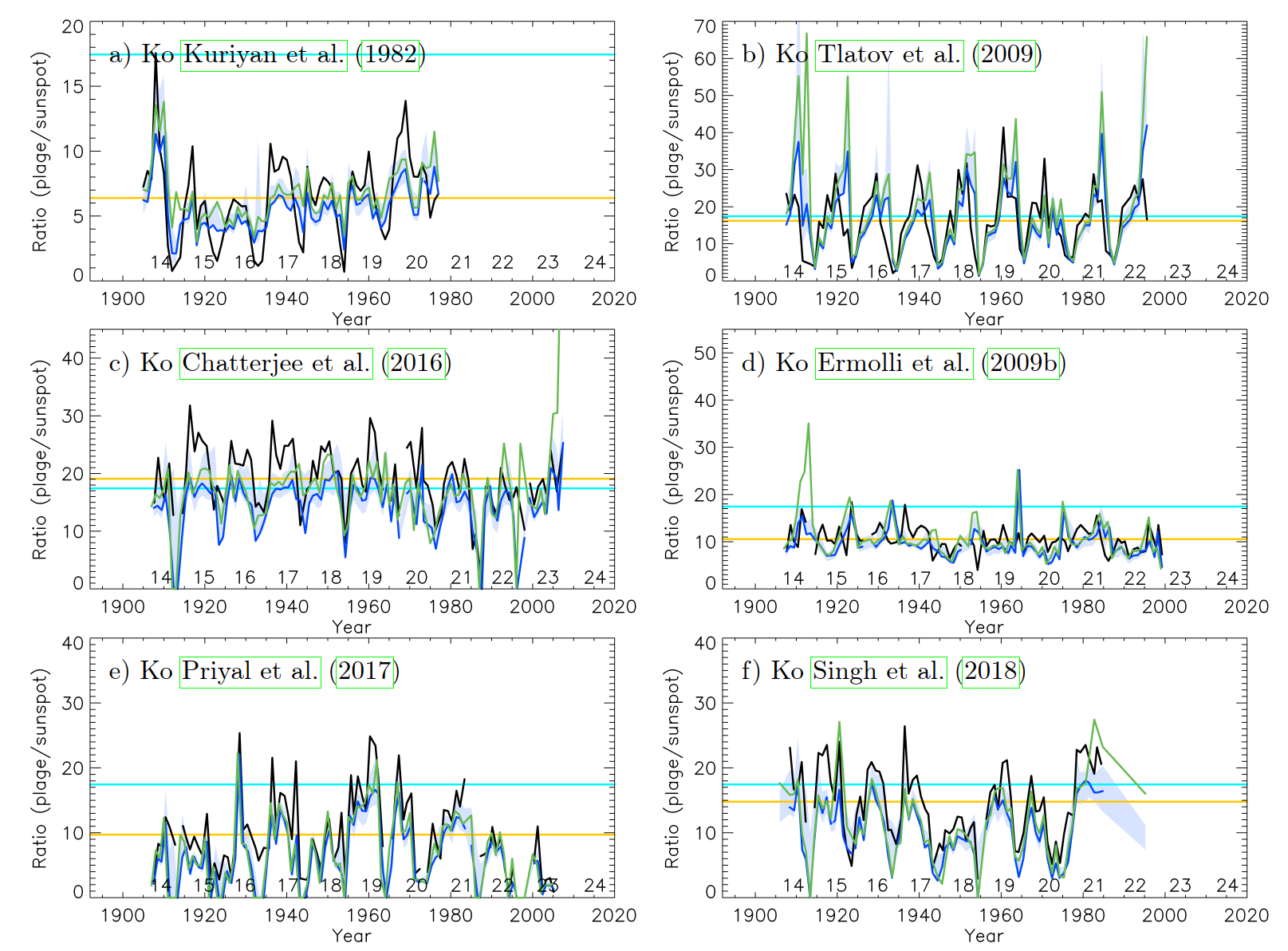}
	\caption{The ratio of the published plage area series to the sunspot area composite by MEA20 (black), as well as the group sunspot number series by CEA17 (blue) and SvSc16 (green). 
	The plage area series shown are those for Ko data by \citet[][a)]{kuriyan_solar_1982}, \citet[][b)]{tlatov_new_2009}, \citet[][c)]{chatterjee_butterfly_2016}, \citet[][d)]{ermolli_comparison_2009}, \citet[][e)]{priyal_long-term_2017}, and \citet[][f)]{singh_variations_2018}. 
	The first two series provide only annual mean values, while the others have daily cadence. The ratios to the group sunspot number series have been multiplied by 2000 to be plotted alongside the ratios to the sunspot areas. The solid lines are annual median values. 
	The blue shaded surface denotes the range of annual values for the ratios from all the sunspot number series used in this study (see Sect. \ref{sec:data}). The ratios for the series with daily values are calculated only for the days for which the plage areas are greater than 0.0005 in disc fraction.  The yellow (light blue) horizontal line represents the average ratio value between the MEA20 sunspot areas and the corresponding plage series (the CEA20 composite plage area series).  The numbers at the lower part of each panel denote the conventional SC numbering and are placed at SC maximum periods. 
	}
	\label{fig:ratiopublishedseriessunspotnumber}
\end{figure*}

\begin{figure*}
	\centering
\includegraphics[width=1\linewidth]{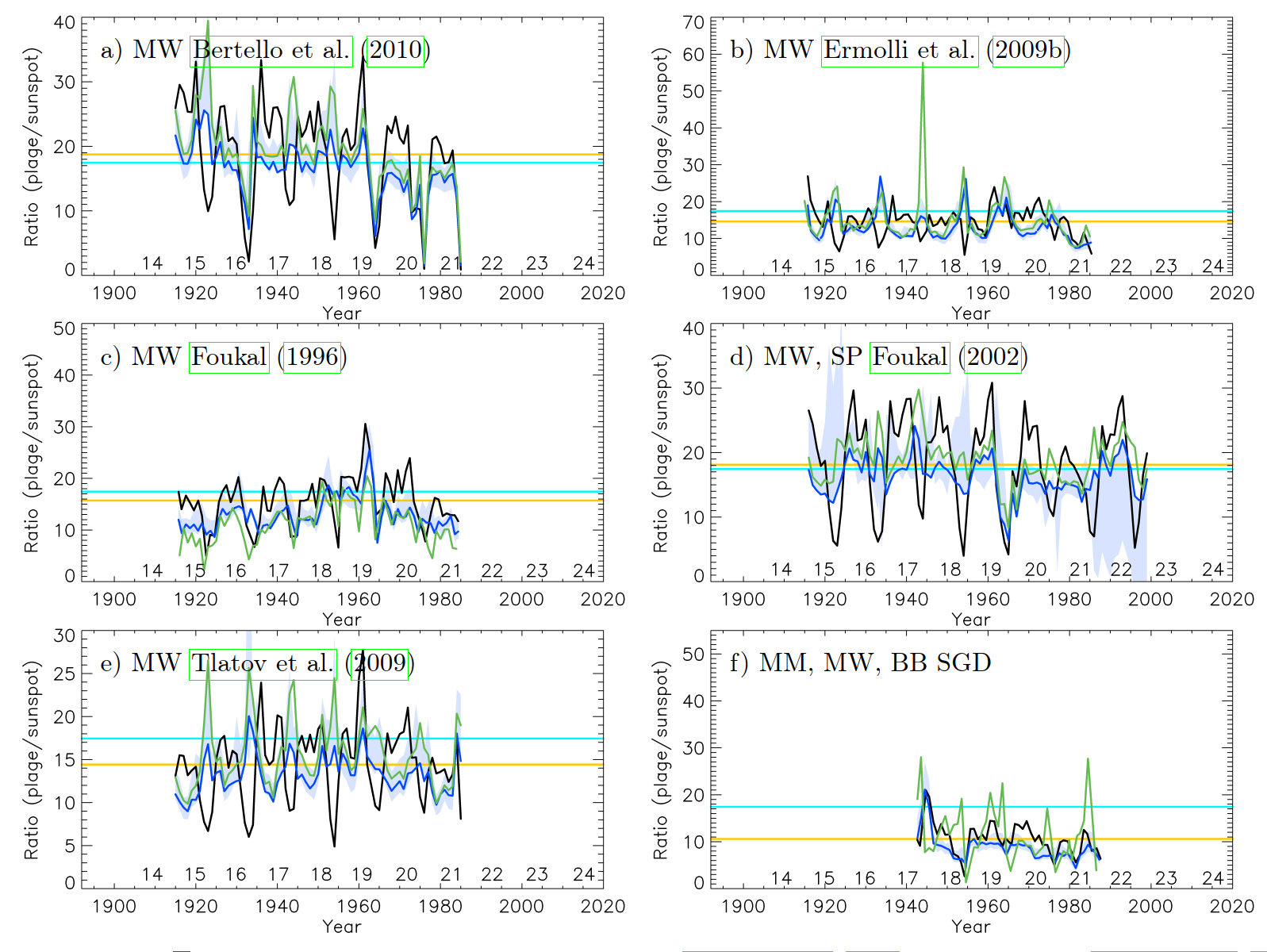}
	\caption{Same as Fig. \ref{fig:ratiopublishedseriessunspotnumber} but for the plage area series for MW data by \citet[][a]{bertello_mount_2010}; MW data by \citet[][b]{ermolli_comparison_2009}; MW data by \citet[][c]{foukal_behavior_1996}; MW and SP data by \citet[][d]{foukal_comparison_2002}; MW data by \citet[][e]{tlatov_new_2009}; and for MM, MW, and BB data by SGD (f);.
	The series in panels b), c) and f) have daily values, while the others have only annual mean values. 
	The \cite{bertello_mount_2010} series, which is a \ca emission index series, was linearly scaled to match the plage areas from our MW data (see Sect. \ref{sec:data}).  }
	\label{fig:ratiopublishedseriessunspotnumber2}
\end{figure*}

To compare our results with those based on earlier published plage area series in a consistent way, we have repeated our analysis by applying the same procedure to such records.
Figures \ref{fig:ratiopublishedseriessunspotnumber} and \ref{fig:ratiopublishedseriessunspotnumber2} show the ratios between the previously published Ko and MW \ca plage area series and the MEA20 sunspot area series.
The figure clearly demonstrates that even when based on the same \ca archive, but processed by different authors using different techniques and digitisation versions, the plage-to-spot ratio exhibits quite different behaviour.

The divergence is yet more substantial when different archives are considered. 
In this respect, we note that five of the considered plage series shown in Fig.~\ref{fig:ratiopublishedseriessunspotnumber} have only annual values and the results derived from them have very different statistics and are thus less informative than those derived from the daily values, while three of those series are in fractional projected areas and the rest in hemispheric areas corrected for foreshortening.
The ratios for the Ko series by \cite{chatterjee_butterfly_2016} and \cite{ermolli_comparison_2009} slightly decrease with time, while for the record by \cite{kuriyan_solar_1982} a slight increase is seen. 
Furthermore, the ratio for the same archive derived from the data by \cite{priyal_long-term_2017} first increases towards SC 19 and then decreases again. 
All MW series show an increase in the ratio over SC 19 and then an abrupt drop over SC 21 consistent with the report of instrumental issues over that period \citep{chatzistergos_analysis_2019}. 
The ratio for the MW series by \cite{foukal_behavior_1996} shows a significant increase over SC 19, suggesting that the plage areas over this period are probably overestimated in this series as also suggested by \citet{ermolli_comparison_2009} and \cite{chatzistergos_analysis_2019}. 
The SGD series returns an average plage to sunspot area ratio of $\sim11$ as compared to the value of 18 obtained from the CEA20 composite. Furthermore, the ratio based on SGD also slightly decreases with time. 
The series by \cite{foukal_comparison_2002} results in a value for the plage to sunspot area ratio similar to ours.
However, in contrast to our result, this ratio decreases over SC 20 and 21 and increases again over SC 22.
This is consistent with the conjecture of a potential problem in the MW data over SC 21 \citep{chatzistergos_analysis_2019}, while the following increase is mainly due to the use of SP data after 1985, when MW stopped operation.
These findings are in agreement with the studies by \cite{ermolli_potential_2018} and \citet{chatzistergos_analysis_2020} who reported inconsistencies within the various published plage area series discussed here.

To our knowledge, only three of the previously published plage area records had been used for irradiance reconstructions: the series by SGD \citep[used e.g. by][]{oster_solar_1982,schatten_importance_1985,foukal_influence_1986,foukal_magnetic_1988,lean_variability_1983,lean_variability_2001}, \citet[][used e.g. by \citealt{foukal_new_2012}]{foukal_comparison_2002}, and \citet[][used by \citealt{ambelu_estimation_2011}]{bertello_mount_2010}. 
All three of these series show a slight decreasing trend with time (Fig. \ref{fig:ratiopublishedseriessunspotnumber}), which is in contrast to our results here. 
This suggests that the competing contributions from sunspots and plage to irradiance variations might have not been accounted correctly in those irradiance reconstructions.
Furthermore, the decreasing trend in the plage-to-spot area ratio with time can potentially disguise any long-term trend in solar irradiance.

\section{Summary and conclusions}
\label{sec:conclusions}
We have studied the relation of plage areas vs. sunspot areas or numbers.
To do so, we used the plage area composite created by \citet{chatzistergos_analysis_2020} by combining the data derived from 38 \ca archives.
For comparison, we have also analysed the relationship using the underlying individual archives.
The sunspot records considered are the sunspot area composite by \citet{mandal_sunspot_2020}, as well as the various available sunspot number and group number series.
The relationship between the daily plage and sunspot areas is best described by a power-law function and remains slightly non-linear also when annual values are considered.
Recently, \citet{nemec_faculae_2022} used a surface flux transport model to show that a
more efficient cancellation of the diffuse magnetic flux in faculae leads to a slower increase of facular coverage with activity compared to spot coverage, eventually resulting in a non-linear relationship between them.
The relationship between the plage areas and the various sunspot number records considered here is also best represented with a power law function, however the linear relation is a good approximation for the annual values.

We also find that the relation between the plage and sunspot areas depends on the bandwidth employed for the observations. 
Furthermore, the relation shows a dependence on the solar cycle strength. 
However, accounting for this dependence when reconstructing plage areas from sunspot areas (numbers) results in only small (minute) improvements.

Furthermore, the relationship between plage and sunspot is affected by the accuracy of the processing of the \ca images. 
We showed that employing various published plage area series from the literature, resulted in rather diverging trends in their ratios to sunspot areas and numbers.
This stands both when considering plage areas derived from data from the same archive (but treated differently) as well as when considering data from different archives.
The advantage of our study is that we have considered not only numerous individual archives but also their composite, which builds on a careful and consistent processing, analysis and cross-calibration of the individual records \citep[see][]{chatzistergos_analysis_2020}.

Thus the results of our study have significant implications for reconstructions of past irradiance variations \citep[e.g.][and references therein]{domingo_solar_2009,yeo_solar_2015,krivova_solar_2018} and stellar activity studies \citep[e.g.][]{lanza_modelling_2003,lanza_stellar_2019,gondoin_contribution_2008,reinhold_transition_2019}. 
Such studies are often carried out assuming a constant ratio of faculae to sunspot areas, which we showed to not be very accurate.

\begin{acknowledgements}
	We thank the observers at the Arcetri, Baikal, Big Bear, Brussels, Catania, Coimbra, Kanzelh\"ohe, Kharkiv, Kodaikanal, Kyoto, Mauna Loa, McMath-Hulbert, Mees, Meudon, Mitaka, Mt Wilson, Pic du Midi, Rome, Sacramento Peak, San Fernando, Schauinsland, Teide, Upice, Vala\v{s}sk\'{e} Mezi\v{r}i\v{c}\'{i}, and Wendelstein sites.
	We thank Sami K. Solanki for helpful discussions during this work.
	We thank the anonymous reviewer for the helpful comments that helped improve this paper.
	We also thank Isabelle Buale for all her efforts to digitise the Meudon
archive, as well Subhamoy Chatterjee, Muthu Priyal, and Andrey Tlatov for providing their published time-series of plage areas.
	T.C. and I.E. thank ISSI for supporting the International Teams 417 "Recalibration of the Sunspot Number Series" and 475 "Modeling Space Weather and Total Solar Irradiance over the Past Century", respectively.
	This work was partly supported 
    by the German Federal Ministry of Education and Research (Project No. 01LG1909C). This research has received funding from the European Union's Horizon 2020 research and innovation program under grant agreement No 824135 (SOLARNET). 
	ChroTel is operated by the Kiepenheuer-Institute for Solar Physics in Freiburg, Germany, at the Spanish Observatorio del Teide, Tenerife, Canary Islands. The ChroTel filtergraph has been developed by the Kiepenheuer-Institute in cooperation with the High Altitude Observatory in Boulder, CO, USA.
	This research has made use of NASA's Astrophysics Data System.
\end{acknowledgements}

\bibliographystyle{aa}
\bibliography{_biblio1}

\appendix

\section{Dependence on cycle strength}
\label{sec:dependenceoncyclestrength}

\begin{table}
	\caption{Cycle averaged values for the various sunspot series.}              
	\label{tab:cycleaveragedvalues}      
	\centering                                      
	\begin{tabular}{l*{4}{c}}          
		\hline\hline                        
		SC&\multicolumn{4}{c}{Sunspot series}\\
				&$\langle s_a\rangle$&\multicolumn{2}{c}{$\langle s_{n}\rangle$}&$\langle s_{gn}\rangle$\\
&MEA20&ISNv1&ISNv2&CEA17\\
\hline 
1&-&-&-& 3.04\\
2&-&-&-& 5.23\\
3&-&-&-& 3.32\\
4&-&-&-& 3.42\\
5&-&-&-& 2.62\\
6&-& 1.06& 1.06& 2.50\\
7&-& 3.24& 3.23& 3.84\\
8&-& 4.78& 4.78& 4.79\\
9&-& 4.43& 4.80& 4.71\\
10&-& 4.04& 4.61& 4.72\\
11& 0.28& 4.43& 4.44& 4.34\\
12& 0.69& 2.84& 2.84& 3.29\\
13&  0.88& 3.25& 3.25& 3.84\\
14&  0.70& 2.69& 2.69& 3.12\\
15&  0.89& 3.67& 3.67& 4.09\\
16&  0.95& 3.41& 3.41& 3.95\\
17&  1.24& 4.80& 4.81& 5.38\\
18&  1.60& 6.22& 5.45& 6.06\\
19&  1.98& 7.60& 6.47& 6.85\\
20&  1.11& 5.10& 4.34& 5.16\\ 
21&  1.73& 6.63& 5.56& 6.24\\
22&  1.66& 6.58& 5.32& 6.04\\
23&  1.32& 4.57& 4.12& 5.01\\
24&  0.88& -& 2.57& -\\
\hline\end{tabular}
	\tablefoot{The values for MEA20 are given in fractional areas multiplied with 1000. Note that the ISNv1 and ISNv2 sunspot number values were divided by 12.08 (ISNv2 values are also multiplied with 0.6) prior to the fit so to bring them roughly to the same level as the group sunspot number series. }
	\end{table}

\begin{figure*}[t]
	\centering
\includegraphics[width=1\linewidth]{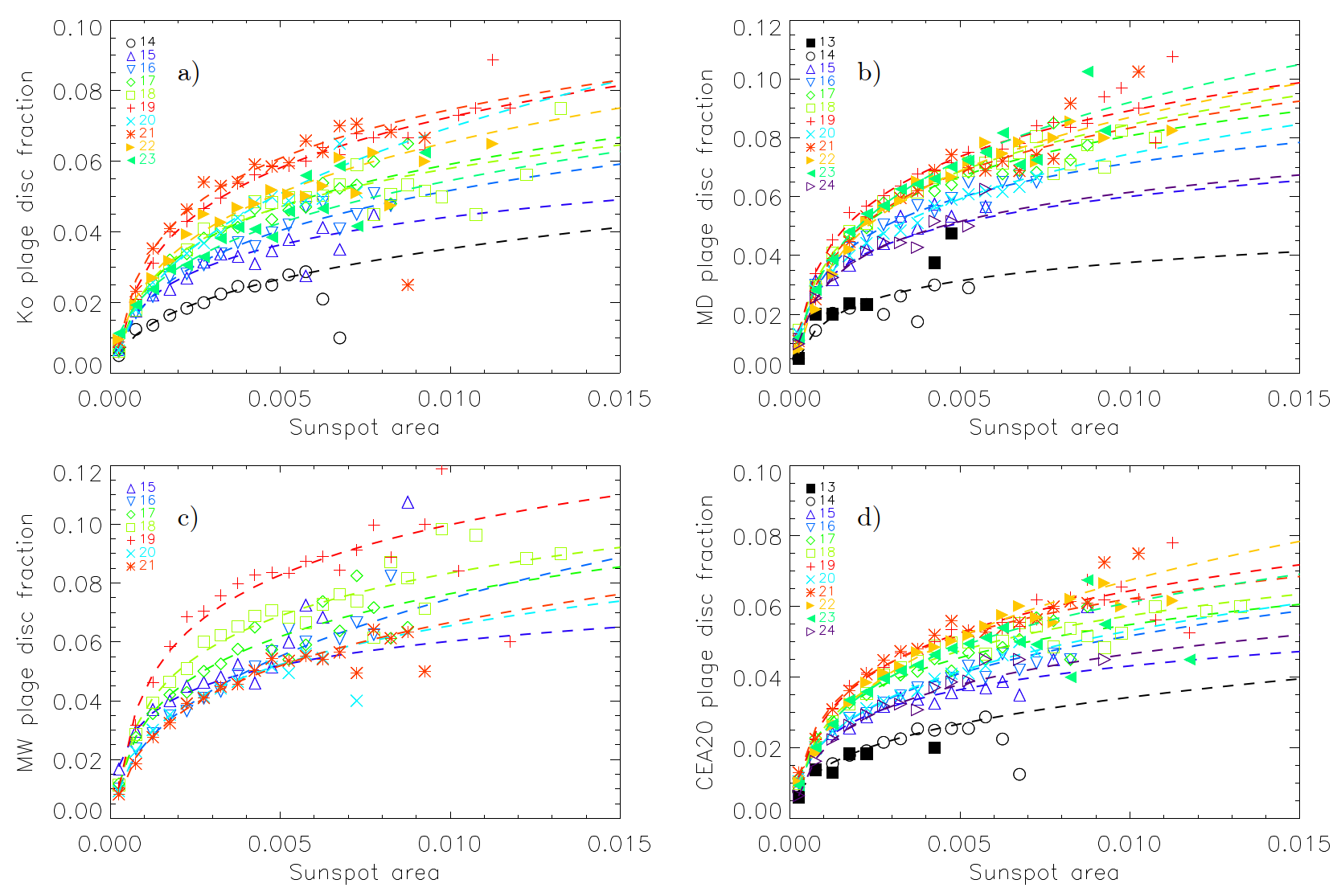}
	\caption{Relation between plage and sunspot areas for individual SC based on the data from the Ko (a), MD (b), MW (c) archives, and the CEA20 plage area composite (d).  The symbols denote the mean values of the PDF matrices (as shown in Fig. 3). The dashed lines are power law  fits and are coloured according to the strength of the SC with the strength increasing from black for the weakest cycle, via dark blue, turquise, green, lime, orange, bright red to dark red for the strongest SC.}
	\label{fig:mwsteepness}
\end{figure*}

\begin{figure*}[t]
	\centering
\includegraphics[width=1\linewidth]{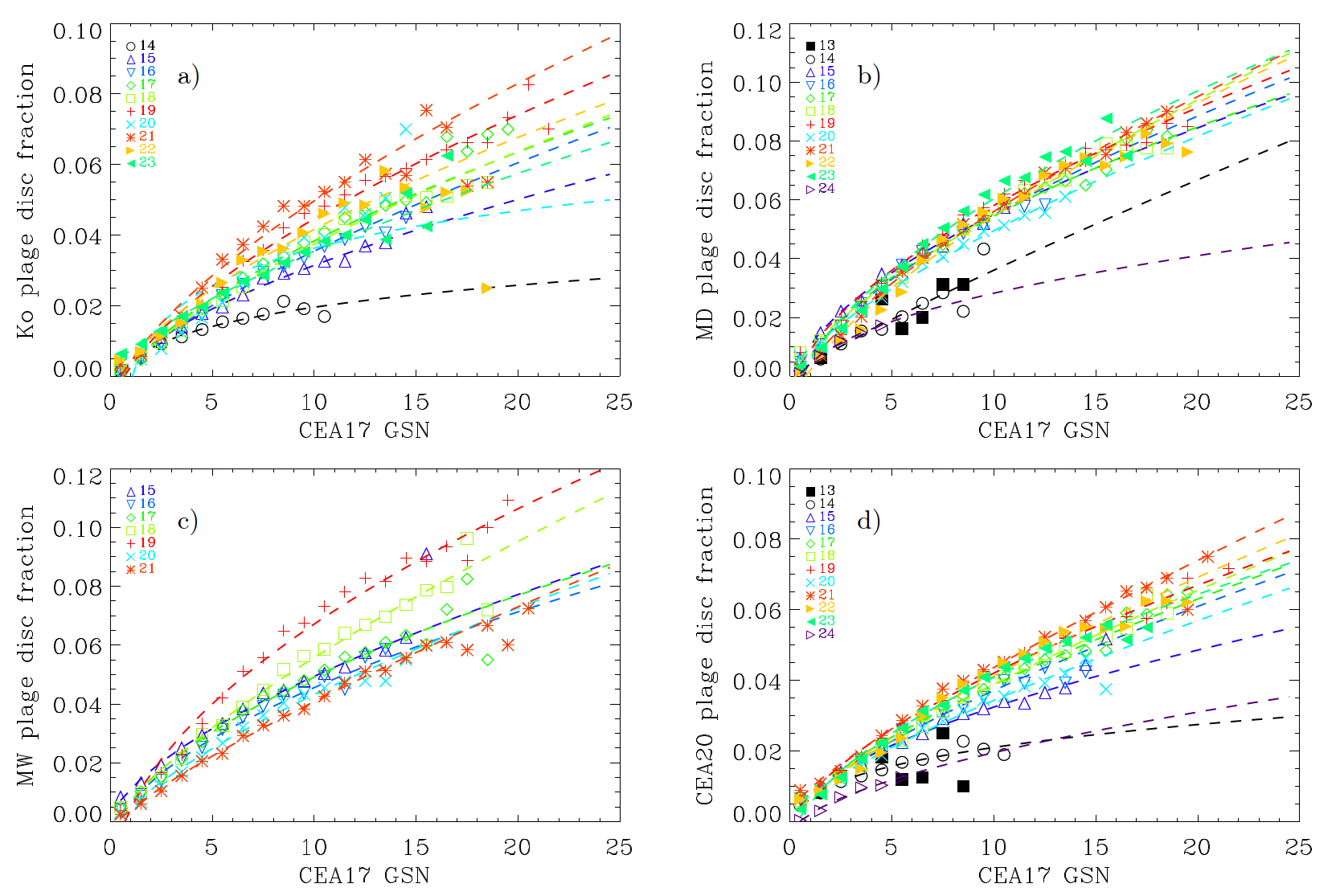}
	\caption{Relation between plage areas and CEA17 GSN for individual SC based on the data from the Ko (a), MD (b), and MW (c) archives as well as the CEA20 plage area composite (d). Shown are the binned values along with power law fits. The lines are coloured according to the strength of the SC with the strength increasing from black for the weakest cycle, via dark blue, turquise, green, lime, orange, bright red to dark red for the strongest SC. }
	\label{fig:mwsteepnessgsn}
\end{figure*}

\begin{figure*}[t]
	\centering
\includegraphics[width=0.8\linewidth]{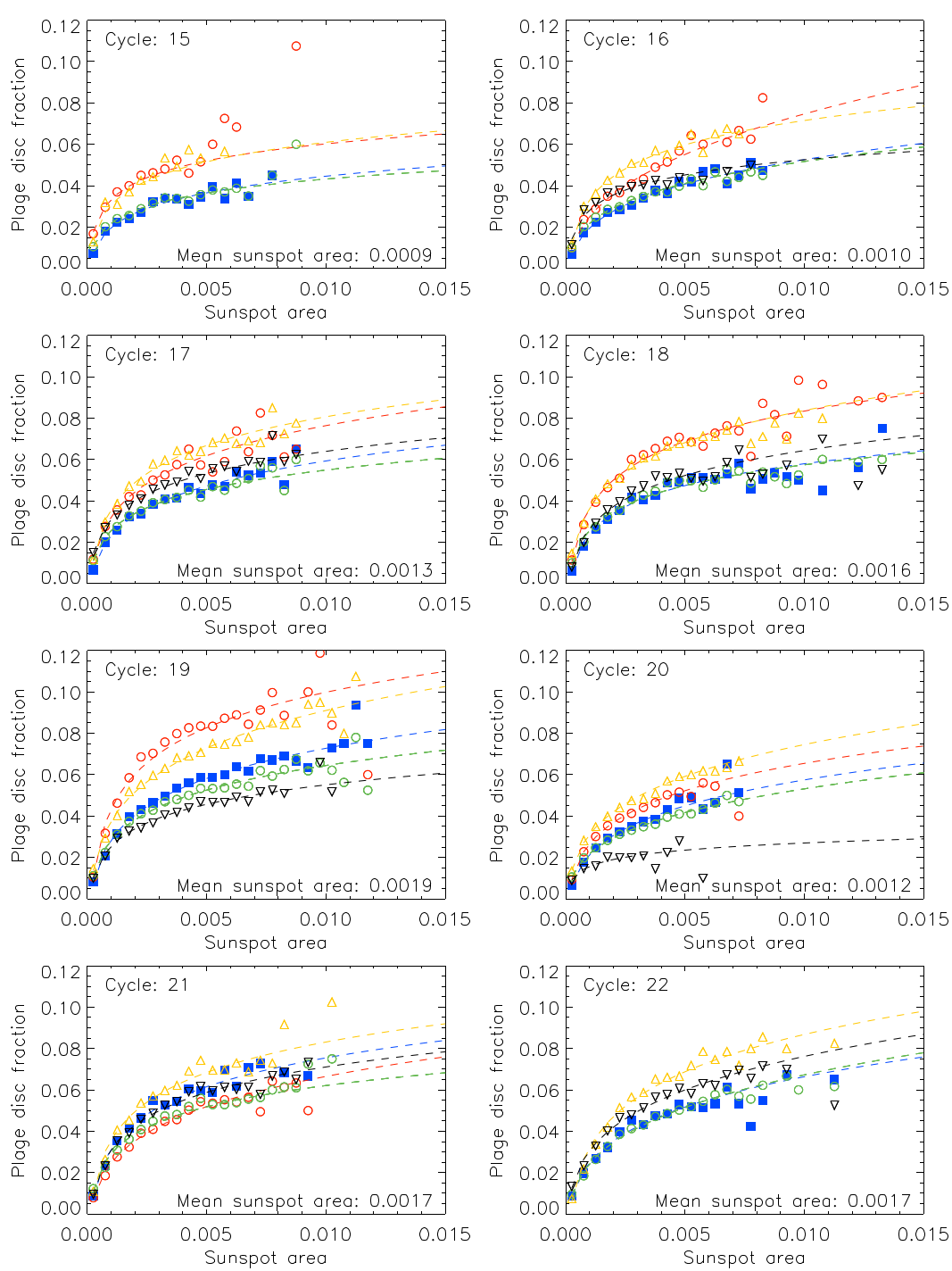}
	\caption{
	Similar to Fig. \ref{fig:mwsteepness}, but now each panel shows the relationships for Co (black downward triangles), Ko (blue squares), MD (orange upward triangles), MW (red circles), and CEA20 composite (green circles) series for individual SCs 15 to 22, as specified in each panel. Also listed in each panel is the cycle-averaged sunspot fractional area.
	}
	\label{fig:steepnessdifcycles}
\end{figure*}

\begin{figure*}
	\centering
\includegraphics[width=1\linewidth]{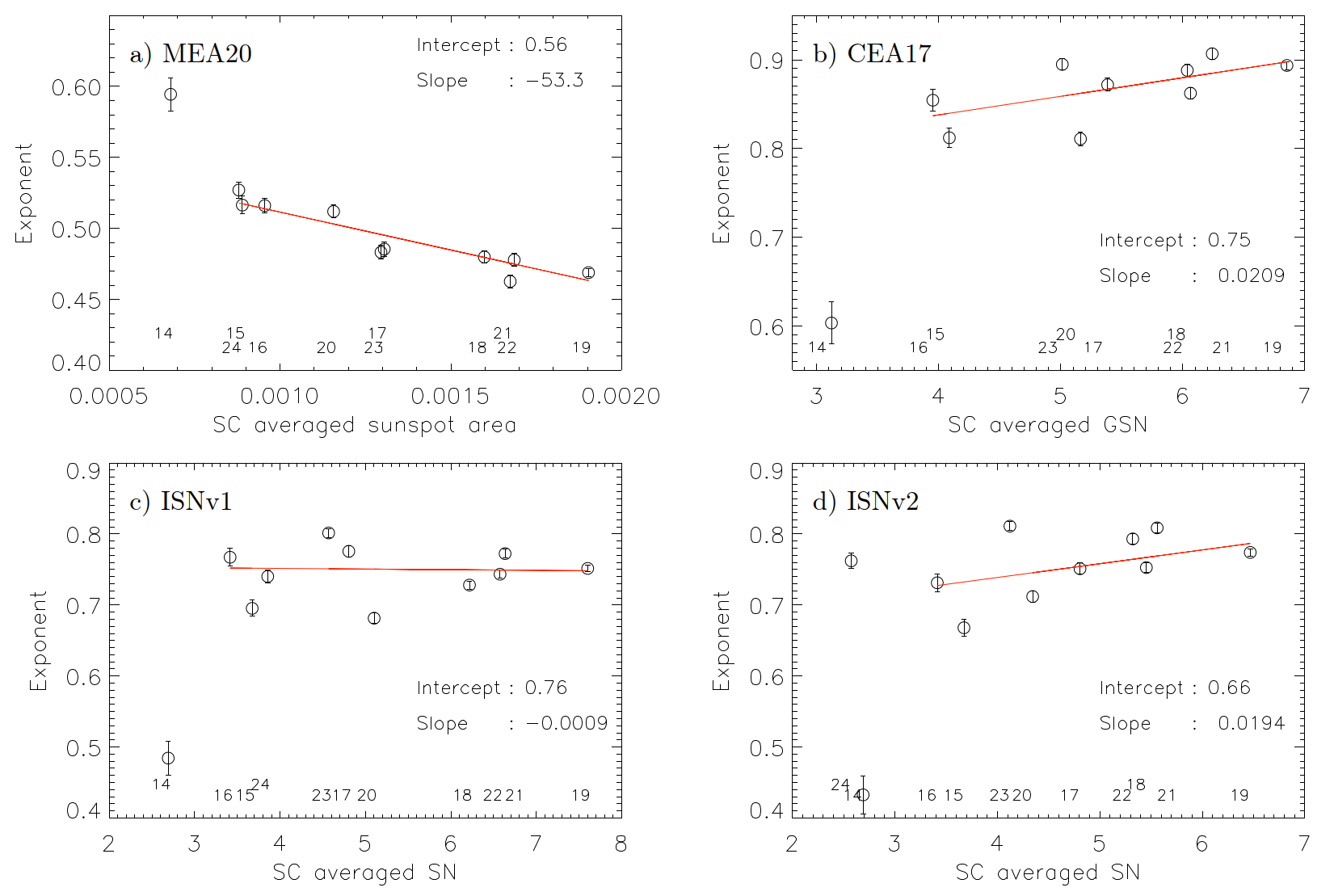}
	\caption{
	Exponents of power law fits to the relation between the CEA20 plage area composite and MEA20 (panel a) sunspot areas as well as CEA17 GSN (panel b), ISNv1 (panel c), and ISNv2 (panel d) series plotted as a function of SC average sunspot number.
	Solid red lines are linear fits to the individual SC values excluding SCs 13 and 14, which do not have enough statistics in the plage area composite. For the comparison to the sunspot number series we also excluded SC 24 since the CEA17 GSN series ends in 2010, while ISNv1 in 2014. The parameters of the linear fits are also given in each panel. Numbers within the panels denote the conventional SC numbering. }
	\label{fig:exponenetsscstrength}
\end{figure*}

Comparing different panels of Figs.~\ref{fig:pdf_indarch_allperiod} and \ref{fig:pdf_indarch_allperiod_cea17}, it is evident that the relationship between the plage areas and the sunspot observations shows some differences among the datasets.
While part of the differences is clearly due to specifics of the individual plage series, there seems to be also a hint at changes in the relationship with time.
Thus, for example, the relationships limited to cycles 22 to 24 (see panels e and f) have a weaker slope than relationships covering earlier periods.
Here we take a closer look at the temporal behaviour of the relationship.

In Fig.~\ref{fig:mwsteepness} we compare the relationship between plage and sunspot areas for individual SCs for the Ko, MD, and MW series, which are the longest series considered in this study, as well as for the CEA20 plage area composite. 
We defined the strength of the SC as the median value of sunspot areas over each SC and use different colours to represent this in Fig.~\ref{fig:mwsteepness}, with dark blue for the weakest SCs through green and yellow for intermediate SCs to dark red for the strongest ones.
Figure \ref{fig:mwsteepness} shows a clear dependence of the plage vs. spot area relationship on the SC strength, with plage areas generally rising faster with sunspot areas for stronger cycles.
This is probably because during stronger cycles, overall more active regions emerge, which in turn leads to an increased number of decaying regions, eventually seeing as faculae. As a result, at a given instant in time there are more facular regions on the surface than there would be during weaker cycles for the same spot coverage.
However, some deviations from strong cycles rising faster are also evident in our results.
Thus, interestingly, the two strongest cycles 19 and 21 show a rather similar relationship (except for MW). 
But in Ko the values for SC 21 are somewhat higher than those for SC19, while the opposite is seen for MD. 
We note, however, that a significant scatter at higher activity levels might affect the fits.
In MW, plage areas for SC 19 reach notably higher values than in other series, while values for SC21 are significantly lower than in other records and than would be expected for such a strong cycle.
This hints at potential problems and inconsistencies in the MW archive over these periods of time.
Figure \ref{fig:mwsteepness}d shows the relation between the composite plage areas from CEA20 and MEA20 sunspot areas for each SC between 13 and 24.
The general form of the studied relation and its dependence on the cycle strength is overall the same as when individual series are considered (panels a--c).
The change with the cycle strength is, however, significantly clearer, with flatter relations for weaker cycles.
Again, SC~19 shows a very similar relationship to the next two cycles in strength, SCs 21 and 22. 

Figure \ref{fig:mwsteepnessgsn} is similar to Fig.~\ref{fig:mwsteepness}, except that it shows the \ca plage area series versus the CEA17 GSN series (instead of the MEA20 sunspot area composite shown in Fig.~\ref{fig:mwsteepness}).
The dependence of the relation between plage areas and the GSN on the SC strength is evident, although less pronounced than between the plage and sunspot areas. 
For most series, it manifests for higher sunspot group number, while it appears to be absent from the MD series.
The results from the archives showing anomalous behaviour over certain SCs when comparing to the sunspot areas exhibit the same behaviour when compared to the GSN series. 
This is unsurprising, considering that the various GSN and sunspot area series are rather consistent with each other over the 20th century.

To highlight the differences between the archives and issues affecting them, Fig.~\ref{fig:steepnessdifcycles} shows the same relationships for the Co, Ko, MD, and MW series, as well as the CEA20 composite, but now for individual cycles 15--22 in each panel.
We note that these four archives employ different nominal bandwidths for their observations, with Co and MD using 0.15\AA, MW 0.2\AA, and Ko 0.5\AA.
Furthermore, the CEA20 composite series was created with RP as the reference dataset, which has a nominal bandwidth of 2.5\AA.
A narrower bandwidth generally results in higher plage areas compared to a broader one \citep{chatzistergos_analysis_2020}.
The effect of the bandwidth on the relation between plage areas and sunspot data is discussed in Appendix \ref{sec:dependencabandwidth}.
The differences between these four archives are generally higher for earlier cycles and are highest for SCs 18 and 19. 
In SC 20, Ko, MD and MW lie comparatively close to each other, whereas Co values are a factor of about 2--3 lower.
The plage areas from the Ko and Co archives are generally close to the composite values and lower than MD and MW areas.
However, in SC 21 MW plage areas are the lowest, while they are highest in SC 19.
Also, in SC19, Co values are lower than all others, while in SC~22 Ko values are clearly the lowest. 
Unfortunately, such a comparison for most of the other plage series is not possible due to their insufficient coverage.
Figure~\ref{fig:steepnessdifcycles} highlights once again that care is needed when using historic observations for past activity studies and that employment of multiple overlapping archives is of great advantage \citep[see also][]{chatzistergos_reconstructing_2021-1}.

To further study the variation of the derived relationship with the SC strength we relate the fit parameters to the SC-averaged sunspot area.
Figure~\ref{fig:exponenetsscstrength} shows the exponents of the power law fits to the relation between plage and sunspot areas 
as a function of the SC-averaged sunspot area.
Also shown (solid red lines) are the linear fits to the values for individual cycles.
Thereby, SCs 13 and 14 were excluded from the fits due to poor data coverage. 
The fits for the CEA17 GSN series also exclude SC 24 since these records do not cover the entire SC 24.
The exponent of the power law fit decreases consistently with increasing SC-averaged sunspot areas with a slope of $\sim-53$. 
The slope of the linear relation between plage areas and the CEA17 GSN increases weakly with SC-averaged GSN with a coefficient of 0.02. 
The opposite behavior of the exponents for the sunspot areas and sunspot number series is because sunspot areas are measured in fractions and are always smaller than one, whereas sunspot numbers are typically greater than 1. 
We note that SC 21 has the lowest (highest) value of the exponent of the fits to sunspot areas (CEA17 GSN).
SC 19 is trailing SC 21, even though SC 19 has a stronger amplitude in all analysed series.
We find similar results for ISNv2 to those reported 
for the CEA17 GSN series with slightly lower slope of 0.019 instead of 0.02.
The results for ISNv1, however show a much lower dependence of the exponents on the cycle strength, with a slope of -9$\times^{-4}$.

Taken the dependence of the relationship on the strength of the solar cycle into account,  Eq. \ref{eq:plf1} and \ref{eq:lnf2} can be written as
\begin{dmath}
    p_a=(0.34\pm0.09)\times s_a^{(0.565\pm0.006-(53.3\pm4.0)\times\langle s_a\rangle)}-(0.004\pm0.008),
    \label{eq:plf1sc}
\end{dmath}
\begin{dmath}
    p_a=(0.005\pm0.001)\times s_{gn}^{(0.75\pm0.03+(0.021\pm0.003)\times\langle s_{gn}\rangle)}+(0.004\pm0.001),
    \label{eq:lnf2sc}
\end{dmath}
where $\langle s_a\rangle$ and $\langle s_{gn}\rangle$ are the cycle averaged sunspot area and group number, respectively.
The values of $\langle s_a\rangle$ and $\langle s_{gn}\rangle$ for the various sunspot series mentioned in this study are given in Table \ref{tab:cycleaveragedvalues}.

\section{Dependence on bandwidth of observation}
\label{sec:dependencabandwidth}

\begin{figure*}
	\centering
\includegraphics[width=1\linewidth]{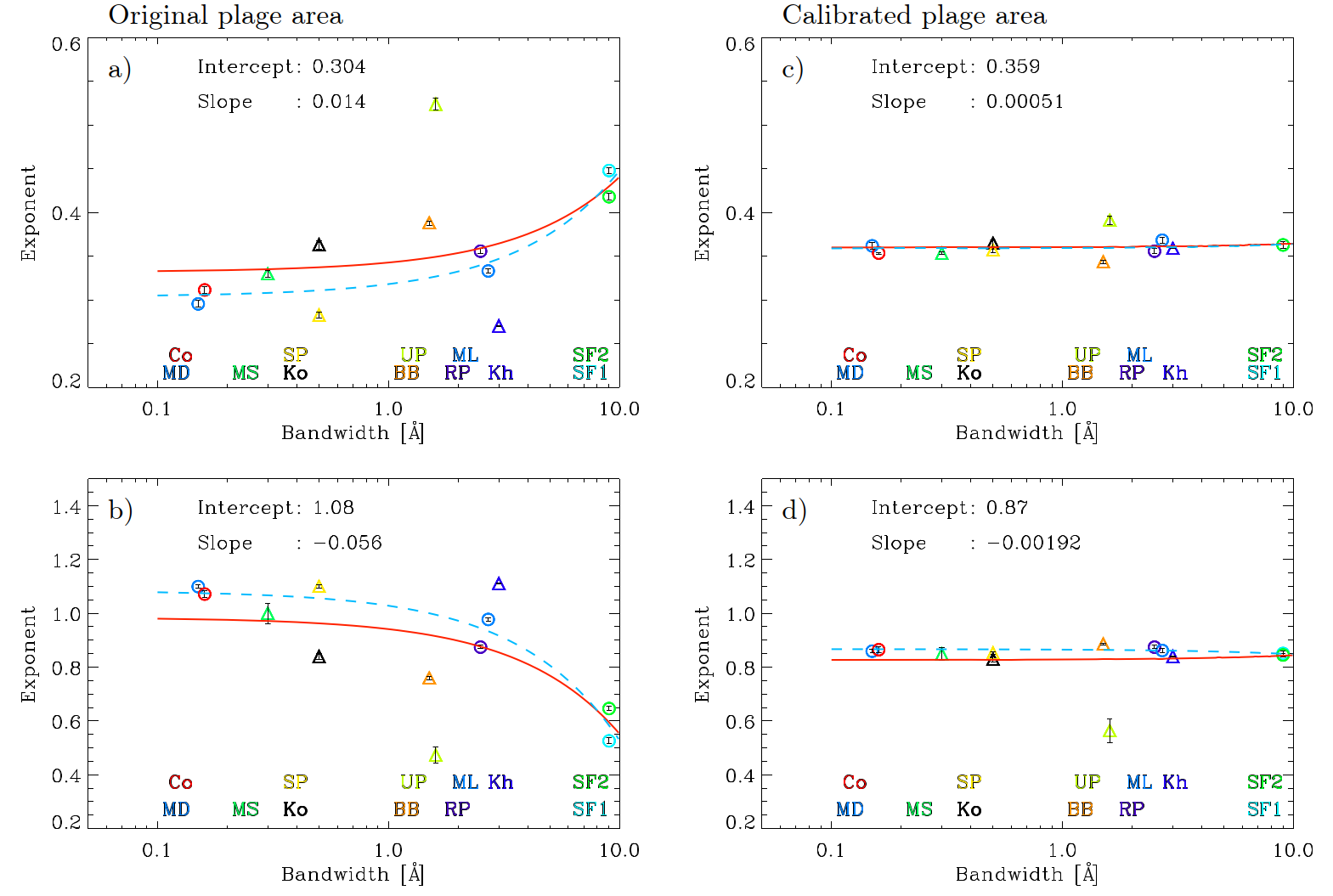}
	\caption{Dependence on bandwidth of the exponent of a power law function and the slope of a linear fit between the plage areas and MEA20 sunspot area (a, c) and the CEA17 GSN (b, d), respectively. Shown are results over SC 23. The archives used to derive the plage areas are BB, Co, Kh, Ko, MD, ML, MS, RP, SF1, SF2, SP, and UP. The results shown in panels a) and b) are for the original plage area series from each archive, while the ones in panels c and d are for the plage areas after the scaling applied by \cite{chatzistergos_analysis_2020} to cross-calibrate all series and merge them into the composite series. The symbols denote the fit result to the mean values of the PDF matrices (as shown in Fig. \ref{fig:pdf_indarch_allperiod}) along with the uncertainty of the fit parameter. Triangles are used for the archives for which there are inconsistencies rendering the bandwidth value uncertain, while circles are used for all other archives. The lines are linear fits; for all points (solid red) and by excluding the archives for which the nominal bandwidth appears inconsistent (dashed blue). The fit parameters of the latter fit are also written in each panel.
	Archive abbreviations at the bottom of each panel follow the order of the respective symbols in the plots.}
	\label{fig:bandwidthcycle23}
\end{figure*}

Here, we discuss how the bandwidth used for the observations affects the relationship between plage areas and sunspot data.
Some systematic differences in the relationship between the plage and sunspot areas among the \ca observations taken with different bandwidths were already hinted by Figs.~\ref{fig:pdf_indarch_allperiod} and  \ref{fig:pdf_indarch_allperiod_cea17}, where the relationship for archives taken with a narrow bandwidth (e.g. MD, MW, or SP) is steeper compared to archives taken with a broader bandwidth (e.g. RP, SF2).

To illustrate this in a more systematic way, Fig. \ref{fig:bandwidthcycle23} (panels a and b) shows the exponent of the power law fits to the average values of the PDF bins (as shown in Figs. \ref{fig:pdf_indarch_allperiod} and \ref{fig:pdf_indarch_allperiod_cea17} , respectively) as a function of the nominal bandwidth of each archive.
To avoid potential effects of the activity variations with time, we only show the data covering the SC 23. We choose this cycle because it is covered by the largest number of the long-running available archives, including also photographic ones. 
There is a slight tendency for the exponent to increase (decrease) with the nominal bandwidth for the sunspot areas (GSN series), although with a high scatter (we remind that the opposite behavior of the exponent is due to the sunspot areas being always lower than one, while sunspot number series are in general greater than 1).
We note, however, that the actual bandwidth of the observations does not always correspond to the nominal one, and considerable ambiguities as well as temporal variations had been noticed for some of the archives  \citep{chatzistergos_analysis_2019,chatzistergos_analysis_2020} hindering a more accurate evaluation of the bandwidth effect.
Such archives are shown by triangles in the figure, and it is evident that exactly these archives \citep[see][]{chatzistergos_analysis_2019,chatzistergos_analysis_2020} are the ones that show the highest scatter with respect to the expected relationships.
For instance, Fig. \ref{fig:bandwidthcycle23} suggests that SP and Kh used a narrower bandwidth, while BB, Ko, and UP a broader bandwidth than their nominal ones.
Furthermore, inconsistencies within the individual archives \citep{chatzistergos_analysis_2019} also affect this comparison, for example the changes in the instrument used to record the BB, ML archives. 

We, further, note that the dependence of the exponent of the fits on the bandwidth of the observations is lifted when using the series after their cross-calibration to RP as done by \cite{chatzistergos_analysis_2020} in order to merge them into the CEA20 plage area composite series. 
This is illustrated in panels c and d of Fig. \ref{fig:bandwidthcycle23}, which are the same as panels a and b, but for the plage area series after applying the scaling by \cite{chatzistergos_analysis_2020}. 
We find the exponents for all cross-calibrated series to lie closer to that of RP compared to the original series (without the cross-calibration). 
The slope of the exponent is significantly reduced compared to the values for the original plage area series.
These results support the corrections done by \cite{chatzistergos_analysis_2020}.

\section{Reconstructing plage areas from sunspot data and different relations}
\label{sec:plagefromsunspot_appendix}

\begin{figure*}
	\centering
\includegraphics[width=1\linewidth]{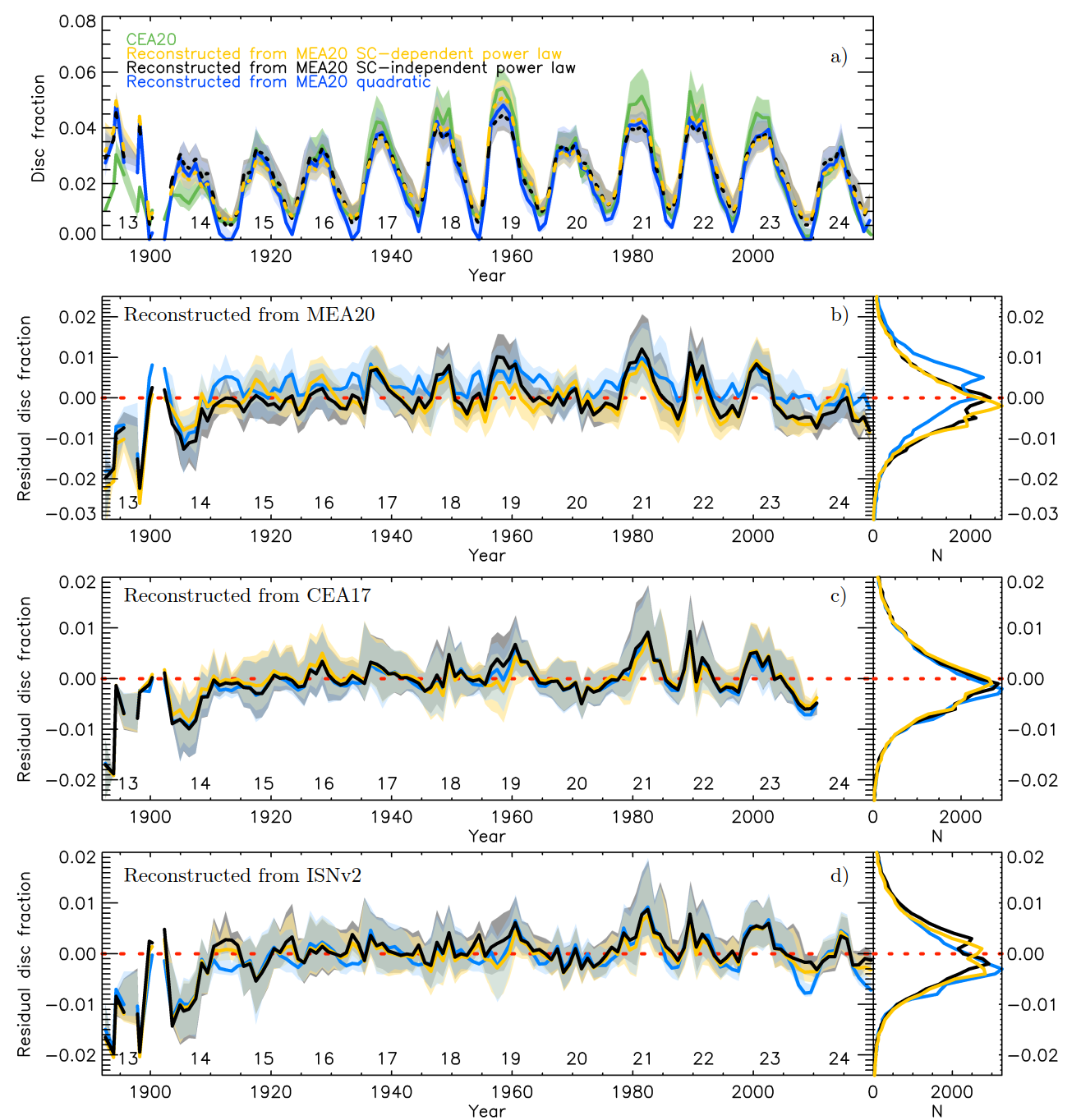}
	\caption{\textit{Panel a: } Reconstructed plage areas from sunspot data along with the CEA20 plage area composite. 	Shown are annual median values (solid lines) along with the asymmetric 1$\sigma$ intervals (shaded surfaces). The numbers below the curves denote the conventional SC numbering and are placed at SC maximum periods. 
	\textit{Panels b--d Left: }Residual areas in fractions of the disc between the CEA20 plage area composite and the plage areas reconstructed from the MEA20 sunspot areas (top panels), the CEA17 GSN series (middle panels), and the ISNv2 series (bottom panels). The reconstructions were produced with a simple linear scaling (blue, only for CEA17 GSN and ISNv2), with a squared root function (blue, only for the MEA20 sunspot areas), with a power law function (black), and a power law function 	with SC-dependent exponents (orange). 
The red dashed line denotes 0 plage area differences. \textit{Panels b--d Right: } Distributions of the differences in bins of 0.001 in fractional areas.}
	\label{fig:residualreconstructedplageseriesfromsunspotnumberseries}
\end{figure*}

Here we use the relationships derived in Sects.~\ref{sec:relations} and Appendix \ref{sec:dependenceoncyclestrength} to reconstruct plage areas from the various sunspot series and analyse the performance of these reconstructions. 
For comparison, we use three different relations: 1) a power law (Eq. \ref{eq:plf1} and \ref{eq:lnf2})  a square-root (linear for CEA17 and ISNv2), both taken to be constant over the entire period analysed in this study (See Table \ref{tab:sgnfitparameters}); and 3) a power law function with a linear dependence of the exponent on the SC-averaged sunspot area (Eq. \ref{eq:plf1sc}) and GSN (Eq.~\ref{eq:lnf2sc}), respectively.

Figure \ref{fig:residualreconstructedplageseriesfromsunspotnumberseries}a shows the reconstructed plage areas from the MEA20 sunspot area series with the three above mentioned relations along with the CEA20 plage area composite. 
Figure \ref{fig:residualreconstructedplageseriesfromsunspotnumberseries}b--d show the residuals between the CEA20 plage area composite and the plage areas reconstructed as described above with all 3 relations from the MEA20 sunspot areas, CEA17 GSN series, and ISNv2, respectively.

In all cases we find the power law relation with the SC-dependent parameters to perform better, however the improvement is rather small.
In particular, the RMS differences between the CEA20 composite series and the series reconstructed with a square-root, a power law, and a power law with SC-dependent exponents used on the MEA20 sunspot area series are 0.0096, 0.0090,  and 0.0086, respectively, while the linear correlation coefficients are 0.83, 0.83, 0.85, respectively.
Hence, the overall improvement when using an SC-dependent power-law function is rather small.
Nevertheless, this reduces the activity dependent effect on the reconstructed plage areas and improves the agreement over cycle maxima, which tend to be slightly underestimated for strong cycles for the reconstructions with a power law with time-independent parameters.

Figure \ref{fig:residualreconstructedplageseriesfromsunspotnumberseries} (panel c) shows the residuals between the CEA20 plage area composite and the reconstructed plage areas derived from the CEA17 GSN record.
Again, for comparison, we also show the plage areas reconstructed using a linear, a power law  with time-independent parameters, and a power law function with SC-dependent exponents.
All three reconstructions perform equally good, with RMS difference to the plage area composite for the linear, power law, and power law with SC-dependent exponents being 0.0079, 0.0079, and 0.0078, respectively, while the linear correlation coefficients are 0.87, 0.88, and 0.88.
We note that using the ISNv2 series for the plage area reconstruction returns a marginally better agreement with the composite than  when CEA17 is employed (Figure \ref{fig:residualreconstructedplageseriesfromsunspotnumberseries} panel c; RMS differences of 0.0076 and 0.0078 for power law function with SC-dependent and -independent exponent and 0.0078 for the linear relationship, all when considering the overlapping periods between ISNv2 and CEA17).
We note that the distribution of residuals of the reconstructed plage areas from ISNv2 to the CEA20 plage area composite exhibit 2 peaks, for instance in the series reconstructed with a power law function one is close to 0 and one at about 0.0055. 
The latter arises due to days with sunspot number in ISNv2 of 0 for which the plage areas are closer to 0.01 (see Fig. \ref{fig:ratiomwsunspotnumberseries}).
Similar double peaks are seen in the residuals for the reconstructions with MEA20 and CEA17, though they are less evident for CEA17.


\end{document}